\documentclass[onecolumn,showpacs,nofootinbib,aps,superscriptaddress, eqsecnum,prd,notitlepage,showkeys,10pt]{revtex4-1}
\usepackage{amssymb}
\usepackage{amsmath}
\usepackage{graphicx}
\usepackage{subfigure}
\usepackage{dcolumn}
\usepackage{natbib}
\usepackage{hyperref}
\newcommand{\ben}{\begin{eqnarray}}
\newcommand{\een}{\end{eqnarray}}
\newcommand{\be}{\begin{equation}}
\newcommand{\ee}{\end{equation}}
\newcommand{\n}{\label}
\newcommand{\no}{\noindent}

\begin{document}
\title{The approach of the three interacting fluids applied to the cosmological constant problem}

\author{M\'onica Forte}
\email{forte.monica@gmail.com}
\affiliation{Departamento de F\'isica, Facultad de ciencias Exactas y Naturales, Universidad de Buenos Aires, 1428 Buenos Aires, Argentina}

\begin{abstract}
We present a cosmological model constituted by three perfect fluids, cold dark matter, vacuum energy and radiation, which interacting with each other lead to an equivalent model of three self-preserved fluids that can be identified with the $\Lambda$CDM model plus a warm dark matter component. The effective energy densities expressed in terms of the global density of energy, its derivatives and interactions, with parameters adjusted with the observational data allow to show the evolution of the vacuum energy. This supports the difference in 120 orders of magnitude of the so-called problem of the cosmological constant and at the same time the strict limits on its density parameter at early times. The best fits parameters of the model, $ H_0= 74.06 $km/sMpc and $q_0=-0.78$, $z_{acc}=0.75$ are concordant with the bibliography and also allow the so-called problems of coincidence and the crisis of age to be alleviated. A geometric analysis performed with statefinders shows the difference with the $\Lambda$CDM model because of the evolution of the effective vacuum density of energy.

\end{abstract}
\maketitle

\section{Introduction}
\label{intro}

Cosmological interactions have been subject of study for a long time, particularly,  the interactions affecting both components of dark sector. The models considered can be devoted exclusively to analyzing the dynamics of both dark components or include any other non interacting component. However, determining the existence of some interaction that also involves radiation or relativistic baryonic matter has aroused interest in relation to the Type Ia supernovae (SNe Ia) observations, that provide the most direct evidence for the current cosmic acceleration and led to the necessity of dark energy (DE). 
Alternative mechanisms contributing to the acceleration evidence or even mimicking the dark energy behavior have been proposed. For example, possible evolutionary effects in SNe Ia events (\cite{Drell:1999dx},\cite{Combes:2003zb});   
 local Hubble bubble (\cite{Zehavi:1998gz},\cite{Conley:2007ng}); modified gravity (\cite{Ishak:2005zs,Kunz:2006ca,Bertschinger:2008zb}), or  unclustered sources of light attenuation (\cite{Aguirre:1999dc},\cite{RowanRobinson:2002iq}), arising in a wide range of well-motivated high-energy physics scenarios, and that could lead to the dimming of SNe Ia brightness (\cite{Avgoustidis:2010ju}).
Also, several authors have recently discussed how the so-called cosmic distance duality (CDD), 
$\frac{D_L}{D_A}(1 + z)^{-2} = 1$ relating the luminosity distance ($D_L$) to the angular diameter distance ($D_A$) of a  given source can be used to verify the existence of exotic physics as well as the presence of systematic errors in SNe Ia observations (\cite{Bassett:2003vu,Uzan:2004my,Holanda:2012at}). In that sense, Holanda et al. \cite{alcaniz} have used recent H(z) measurements from passively evolving galaxies to obtain cosmological model-independent distance modulus and impose constraints on cosmic opacity by comparing these data with  the Union2 and the SDSS compilations. Both, Union2 and SDSS (SALT2) compilations are in full agreement with a perfect transparent and flat universe whereas the SDSS compilation that uses MLCS2K2 light-curve fitting rules out such a possibility by $\sim 3.5\sigma$. In this regard, a possible coupling of photons to particles beyond the standard model of particle physics modifying the apparent luminosity of sources has recently been considered (for a review see \cite{Jaeckel:2010ni}). In this paper we will put aside these differences, adjusting the model parameters with the Hubble function in the study of a system of three interacting components that does not appeal to dark matter (DM) or dark energy but is sourced by a vacuum energy, baryonic nonrelativistic  matter and radiation. It will show the compatibility of this interactive model
with a model similar to $\Lambda$CDM, where dark energy will be represented by a density of energy of vacuum while the dark matter component will have two inputs: a cold dust type fluid (CDM) that includes baryonic matter and another novel type of DM sometimes called warm dark matter (WDM),  with a slightly positive pressure \cite{Harko:2011kw,Wei:2013et,Perivolaropoulos:2008ud,Perivolaropoulos:2011hp,deVega:2011si}.


\section{The three interactive fluids system}


We consider a spatially flat homogeneous and isotropic universe described by the Friedmann-Lemaître-Robertson-Walker (FLRW) metric with line element given by $ds^2 = -dt^2 + a^2(t)(dx^2 + dy^2 + dz^2)$ 
being $a(t)$ the scale factor. Our model of universe is filled with three interacting components that describe dust, vacuum energy and radiation fluid. They have energy densities $\rho_1$, $\rho_2$, and $\rho_3$ and pressures $p_1$, $p_2$, and $p_3$ respectively, so that the evolution of the FLRW universe is governed by the Friedmann and conservation equations,
\be
\n{01}
3H^2=\rho=\rho_1+\rho_2+\rho_3, 
\ee

\be
\n{02}
\dot\rho_1+\dot\rho_2+\dot\rho_3+3H(\rho_1+\rho_2+\rho_3+p_1+p_2+p_3)=0, 
\ee

\no where $H = \dot a/a$ is the Hubble expansion rate. 
Eq.(\ref{02}) describes a mix of three interacting fluids with bare equations of states (EoS) $\omega_i=p_i/\rho_i$, for $p_1=0$, $p_2=-\rho_2$ and $p_3=\rho_3/3$. Then, $\rho_1$ represents a DM component, $\rho_2$ plays the role of energy of vacuum  and $\rho_3$ can be associated with a radiation term. At this point, we introduce the three interaction terms $3HQ_1$, $3HQ_2$ and $3HQ_3$, so that the conservation equation Eq.(\ref{02}) is split into three balance equations
\be
\n{03}
\rho'_1+\rho_1=Q_1, 
\ee
\be
\n{04}
\rho'_2=Q_2, 
\ee
\be
\n{05}
\rho'_3+\frac{4}{3}\rho_3=Q_3, 
\ee

\no where the interaction terms satisfy the condition
\be
\n{06}
Q_1+Q_2+Q_3=0, 
\ee

\no to recover the whole conservation equation

\be
\n{07}
\rho'=-\rho_1-\frac{4}{3}\rho_3,
\ee
\no and ' stands for derivatives with respect to the variable $\eta=\ln(a/a_0)^3$.

 After differentiating the Eq.(\ref{07}) and using Eqs. (\ref{03}) - (\ref{05}), we obtain 
\be
\n{08}
\rho''=\rho_1+\frac{16}{9}\rho_3 - Q_1 - \frac{4}{3} Q_3.
\ee

Then, using (\ref{01})and  (\ref{03})-(\ref{08}), we can describe the $(\rho_1,\rho_2,\rho_3)$ variables as functions of $(\rho,\rho',\rho'')$ variables and the coupling functions, $Q_1$ and $Q_2$,

\be
\n{09}
\rho_1= -[4\rho'+3\rho''-4Q_2-Q_1], 
\ee\be
\n{10}
\rho_2= \frac{1}{4}[4\rho+7\rho'+3\rho''-4Q_2-Q_1], 
\ee
\be
\n{11}
\rho_3= \frac{3}{4}[3\rho'+3\rho''-4Q_2-Q_1]. 
\ee

The third order differential equation  for the total density  of  energy is obtained by differentiation the Eq.(\ref{08}) and following the Ref. \cite{Chimento:2009hj} we get 
\be
\n{12}
3\rho''' + 7\rho'' + 4\rho' = 4(Q_2+Q'_2) + Q'_1.
\ee

Thus, once the interactions $Q_i=Q_i(\rho,\rho',\rho'')$ are specified, we obtain the density of energy $\rho$ by solving the source equation (\ref{12}), whereas the component energy densities  $\rho_1$, $\rho_2$, and $\rho_3$ are obtained after inserting $\rho$, $\rho'$ and $ \rho''$ into Eqs.(\ref{09})-(\ref{11}).


\section{The non-transversal interactions and the equivalent non interactive model}


In order to get some insight on the nature of the model, we analyze the set of interactions 

\begin{eqnarray}
\nonumber
Q_1&=&\mu(\rho-\rho''),\\
\n{13}
Q_2&=&\alpha\rho',\\
\nonumber
Q_3&=&-(\mu\rho+\alpha\rho'-\mu\rho''),
\end{eqnarray}

\no that do not satisfy the relationship $\sum_{i=1}^3(1+\omega_i)Q_i= 0$, nicknamed transversal, and are linearly dependent on $\rho$, $\rho'$ and $ \rho''$.
 
The choice (\ref{13}) produces a source equation for the total density of energy $\rho$
\be
\n{14}
\rho'''+ \mathcal{A}\rho'' +\mathcal{B} \rho'=0,
\ee
where $\mathcal{A}=(7-4\alpha)/(3+\mu)$ and $\mathcal{B}=(4-4\alpha-\mu)/(3+\mu)=\mathcal{A}-1$.

If the intention is to study the problem of the value of constant cosmological density of energy, these choices are obligatory in order to obtain a component of the constant cosmological type, that is, we must discard a term proportional to $\rho$ in the master equation (\ref{14}). On the other hand, if we assume that at early times the model is composed mainly with energy of vacuum, it is possible to consider that its variation is fundamentally the variation of global energy, that is $Q_2 \sim \rho'$. At the same time, the generation of DM particles must be proportional to the available energy  $Q_1\sim\rho$ and also related to the slowdown of the decrease in global density, that is, $ Q_1 \sim -\rho''$

The equation (\ref{14}) have solutions $\rho \sim e^{-\lambda\eta}\equiv (1+z)^{3\lambda}$  whose exponents are the three roots $\lambda_1=0$, $\lambda_{\pm}= \left(\mathcal{A}\pm\sqrt{\mathcal{A}^2-4\mathcal{B})}\right)/2$, of the secular equation $\lambda^3-\mathcal{A}\lambda^2+\mathcal{B}\lambda=0$. The general solution of equation (\ref{14})  as a function of redshift is
\be
\n{15}
3H^2=3H_0^2\left[b_{\Lambda}+b_{Dust}(1+z)^{3}+b_{Warm}(1+z)^{3\mathcal{B}}\right],
\ee
where we have fixed $\lambda_-=1$ to recover the contribution of non-relativistic matter at early times, and so  it results in that $\lambda_+=\mathcal{B}$. In this case are $\mathcal{A}-2>0$ and $\mathcal{B}>1$. The constants $b_i$ satisfy the condition $b_{\Lambda}+b_{Dust}+b_{Warm}=1$ in order to be compatible with flat universe.

Among the parameters  ($b_{\Lambda}$,$b_{Dust}$,$\mathcal{B}$,$H_0$), the only one who can take negative values is $b_{\Lambda}$ and it plays the role of a cosmological constant. This sign is admissible if, as it is argued in \cite{carroll}, the vacuum energy is non-dynamical so that a negative value cannot induce any instabilities. In that sense, negative cosmological constant has attracted a lot of attention in brane scenarios \cite{Randall:1999vf,Sahni:2002dx,Chimento:2009rw} and also due to the AdS/CFT correspondence,\cite{Cvetic:2003zy}. However, there are important effects in gravitational thermodynamics, as the Antonov's gravothermal instability, demonstrating a positive cosmological constant is nowadays the best candidate for dark energy \cite{Axenides:2012bf}. Here, we only consider $b_{\Lambda}>0$. 

\begin{figure}[hbpt]
  \centering
  \subfigure[]{\includegraphics[height=5cm,width=0.45\textwidth]{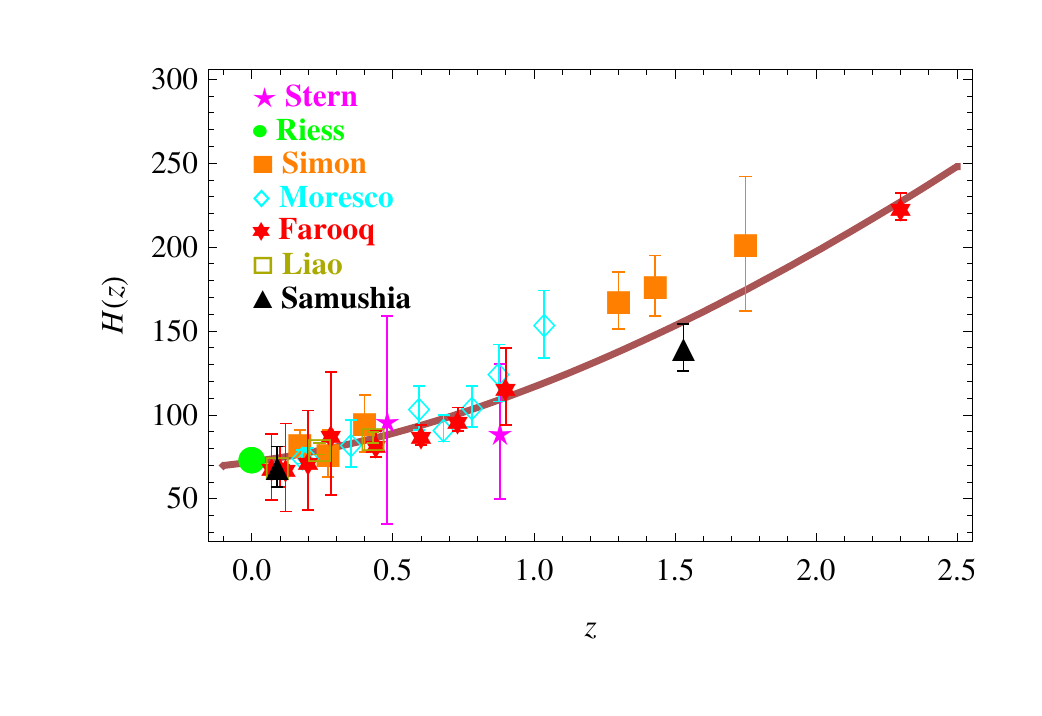}
\label{Fig:. Figura1a}}
  \subfigure[]{\includegraphics[height=6cm,width=0.50\textwidth]{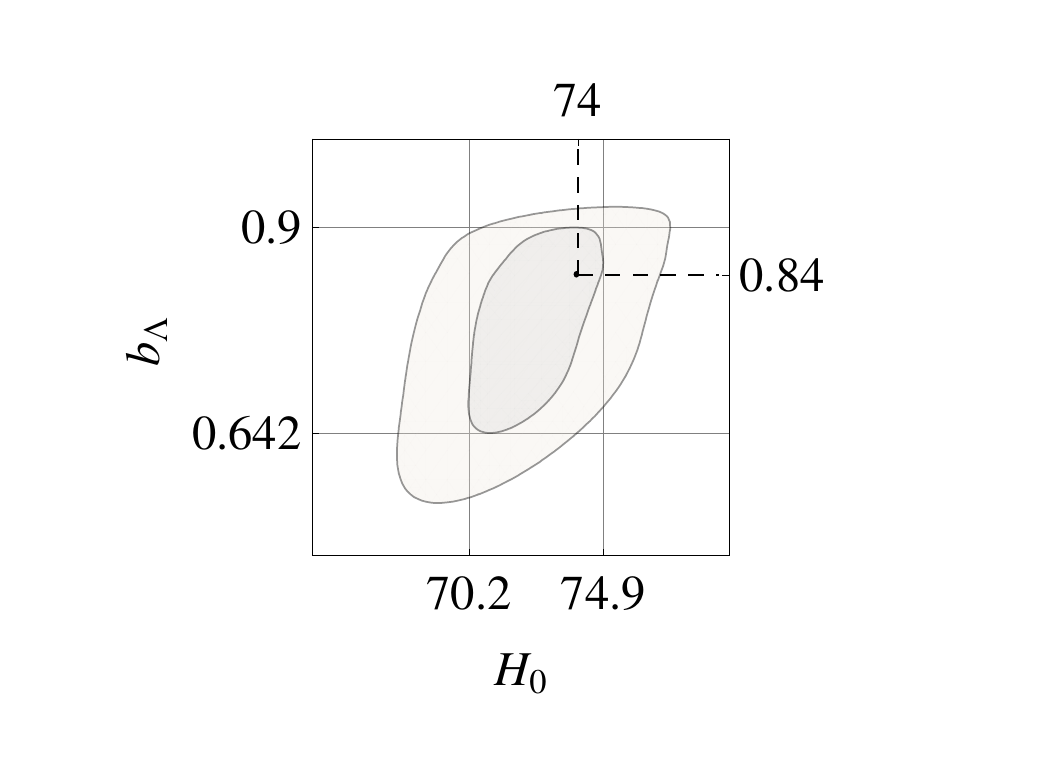}
\label{Fig:. Figura1b}}
  \caption{
a)\scriptsize{Hubble curve (\ref{09}) of the equivalent model with the best fit $H_0=74 \mathrm{km/sMpc}$, $b_{\Lambda} = 0.843$, $b_{Dust}=0.05$ and $\mathcal{B}=1.242$ and all the different Hubble data from \cite{Stern:2009ep,Riess:2009pu,Simon:2004tf,Moresco:2012by,Farooq : 2013 hq,Liao : 2012 bg,Samushia:2006fx}},\\
b)\scriptsize{Marginalized bi-dimensional confidence regions $1\sigma$ and $2\sigma$ in the parameter space $H_0$  vs $b_{\Lambda}$ for the equivalent model. The best fit parameter $b_{\Lambda}=0.843^{+0.06}_{-0.20}$ can be considered as the density of energy of a cosmological constant dark energy  $\rho_{DE}$, in units of $H_0^2$, and the best fit value $H_0=74^{+0.9}_{-3.8} \mathrm{km/sMpc}$ as the values of the actual Hubble factor usually recorded in the literature.}}
  \label{Fig:. Figura1ab}
\end{figure}

To identify the components of the global equivalent model (\ref{15}), we used the Hubble function H(z) method to adjust the parameters $(b_{\Lambda}, b_{Dust}, \mathcal{B},H_0)$ \cite{Press}.
 This method of constraint parameters appeared to be more suitable than the use of SN Ia because SNe Ia observations are affected by at least four different sources of opacity (the Milky Way, the hosting galaxy, intervening galaxies, and the Intergalactic Medium). Instead, the current H(z) measurements are obtained from ages estimates of old passively evolving galaxies, which relies only on the detailed shape of the galaxy spectra, not on the galaxy luminosity. Therefore, differently from $D_L$ measurements from SNe Ia, H(z) observations are not affected by cosmic opacity  since this quantity is assumed to be not strongly wavelength dependent on the optical band (see Avgoustidis et al. 2009 and references therein for more details). We used the data base from Moresco, Verde et all \cite{Moresco:2012by},  and the adjustment gave the best fit results set $H_0=74 \mathrm{km/sMpc}$, $b_{\Lambda} = 0.844$, $b_{Dust}=0.047$ and $\mathcal{B}=1.243$, with $\chi^2_{min}=13.8834$, that is, a good value per degree of freedom $\chi^2_{dof}=0.992$.  
The Fig.:\ref{Fig:. Figura1a} exhibits a very good adjust of the curve (\ref{09}) of H(z) for the best fit parameters, with all the data about the Hubble function \cite{Stern:2009ep,Riess:2009pu,Simon:2004tf,Moresco:2012by,Farooq : 2013 hq,Liao : 2012 bg,Samushia:2006fx}.

\begin{figure}[hbpt]
  \centering
  \subfigure[]{\includegraphics[height=5cm,width=0.4\textwidth]{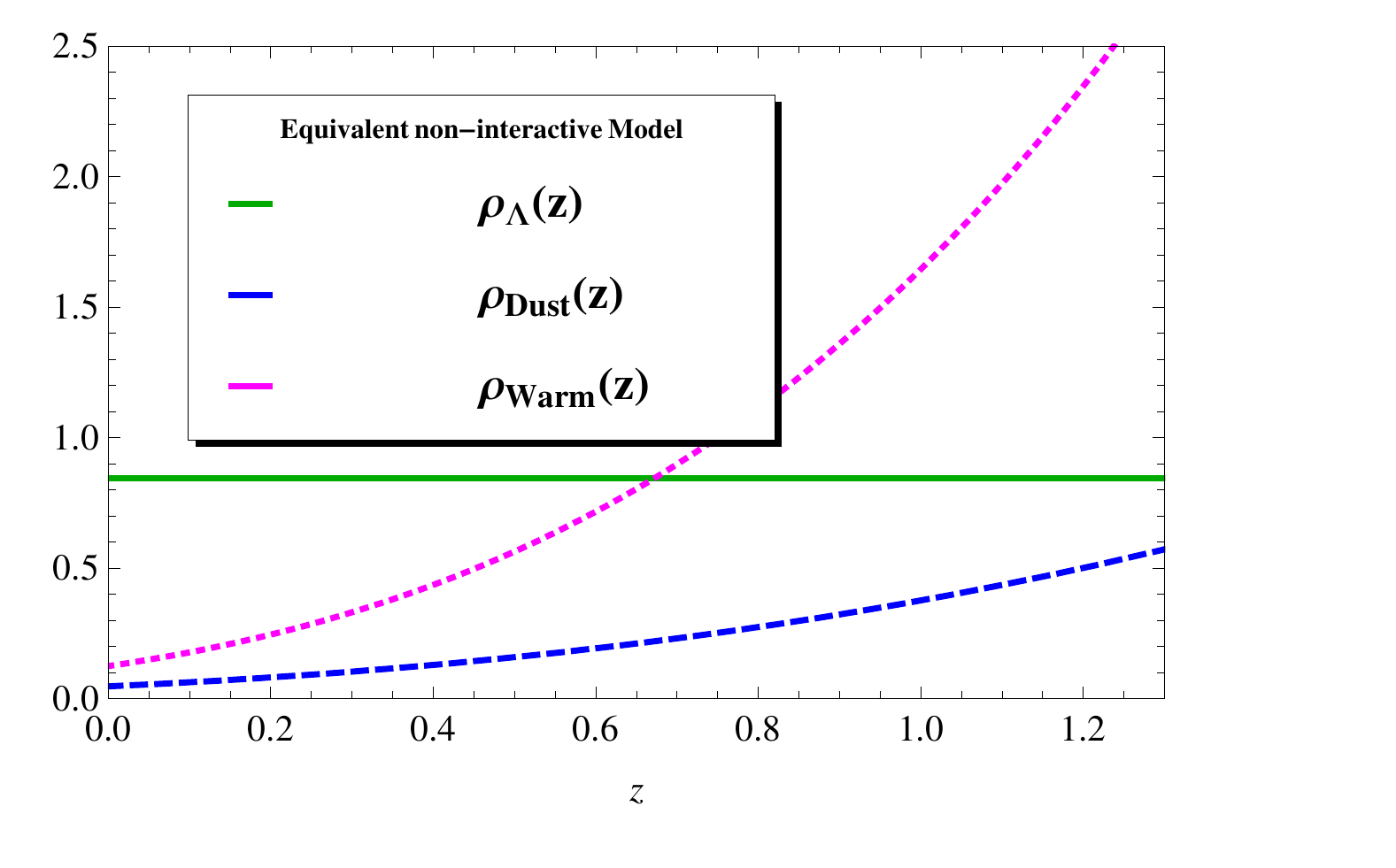}
\label{Fig:. Figura2a}}
  \subfigure[]{\includegraphics[height=6cm,width=0.55\textwidth]{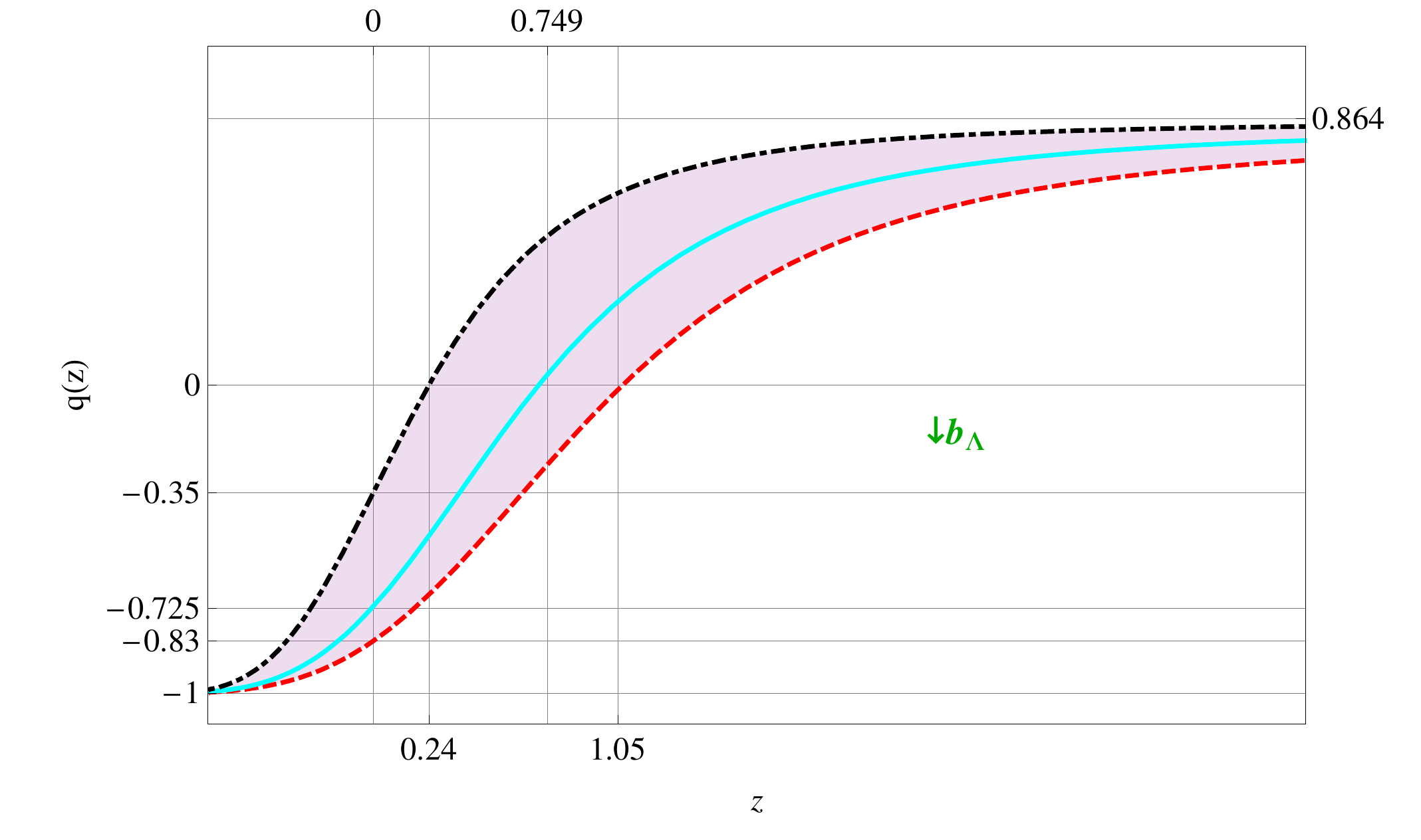}
\label{Fig:. Figura2b}} 
  \caption{a)\scriptsize{Behavior of the densities of energy of the equivalent model. The three components are identified with a cosmological constant dark energy  $\rho_{DE}=\rho_{\Lambda}$,  baryonic non relativistic matter $\rho_{baryonic}$  and  dark matter $\rho_{DM}$, in units of $3H_0^2$ for the best fit parameters $b_{\Lambda}=0.843$, $b_{Dust}=0.047$ and $\mathcal{B}=1.243$.},\\
b)\scriptsize{Approximated $1\sigma$ region for the evolution of deceleration parameter $q(z)$ obtained in the equivalent model for the $1\sigma$ range of parameter $b_{\Lambda}$. The solid cyan curve corresponds to the best fit parameter, $b_{\Lambda}=0.834$, while values maximum and minimum of the $1\sigma$ range of $b_{\Lambda}$, decreases (dashed red curve) and increases (dot-dashed black curve), respectively, the values of the deceleration parameter. The variation is negligible in the distant past but is very important in the transition stage where the beginning of the acceleration occurs much earlier for the highest value $b_{\Lambda}=0.9$. For the three curves we used $H_0=74 \mathrm{km/sMpc}$, $b_{Dust}=0.047$ and $\mathcal{B}=1.243$.}}
  \label{Fig:. Figura2ab}
\end{figure}

The Fig.:\ref{Fig:. Figura1b} shows the marginalized bi-dimensional confidence regions $1\sigma$ and $2\sigma$ in the parameter space $b_{\Lambda}$ vs $H_0$ for the equivalent model (\ref{15}). The best fit parameters $b_{\Lambda}=0.843^{+0.06}_{-0.20}$ and $H_0=74^{+0.9}_{-3.8} \mathrm{km/sMpc}$ are compatible with the density of energy of a cosmological constant dark energy $\rho_{DE}=\rho_{\Lambda}$ and the values of the actual Hubble factor usually recorded in the literature. 
Note that the best fit values above allow us to identify $b_{Dust}=0.047$ as the density of energy of baryonic matter and the third contribution in (\ref{15}) can be regarded as a dark matter component with a small but non zero pressure, called warm dark matter (WDM).  This latter issue was studied in \cite{Harko:2011kw} where it is pointed out that the necessity of considering non-standard dark matter models with pressure is justified by the uncertainties in the knowledge about the nature of the dark matter particles, as well as by the fact that these models give a much better description of the observational results, as compared to the pressureless case.
For the above best fit parameter values, the Fig.:\ref{Fig:. Figura2a} shows the densities of energy of the non interacting equivalent components, which satisfy the equation (\ref{15}), where the fluids can be identified, at least at the present epoch, as a cosmological constant dark energy, baryonic non relativistic matter  and dark matter with small but non zero pressure (WDM). The Fig.:\ref{Fig:. Figura2b} shows the approximated $1\sigma$ region for the evolution of deceleration parameter $q(z)=-1-3\rho'/2\rho$ obtained in the equivalent model for the $1\sigma$ range of parameter $b_{\Lambda}$, using the best fits $H_0=74 \mathrm{km/sMpc}$, $b_{Dust}=0.047$ and $\mathcal{B}=1.243$.The solid cyan curve corresponds to the best fit parameter, $b_{\Lambda}=0.834$, while values maximum and minimum of the $1\sigma$ range of $b_{\Lambda}$, decreases (dashed red curve) and increases (dot-dashed black curve), respectively, the values of the deceleration parameter. The variation is negligible in the distant past but is very important in the transition stage where the beginning of the acceleration occurs much earlier when considering the highest value $b_{\Lambda}=0.9$.

At this point, the equivalent model can be regarded as a $\Lambda$WDM model, assuming that the pressure of DM can be  small, but nonzero.  It is claimed that WDM can successfully reproduce
the astronomical observations over all the scales (from small/galactic to large/cosmological scales)\cite{deVega:2011si}.


\section{The three effective interactive fluids}

The effective model is constituted by interactive fluids whose densities of energy $\rho_i$, for the set (\ref{13}) of interactions $Q_i$, can be written as functions of the redshift as

\begin{subequations}
\n{16}
\be
\n{16a}
\frac{\rho_1}{3H_0^2}=\mu b_{\Lambda}+(1-4\alpha)b_{Dust}(1+z)^{3} + \mu (1+\mathcal{B})b_{Warm}(1+z)^{3\mathcal{B}}\ \ \ \ \ \ \ 
\ee
\be
\n{16b}
\frac{\rho_2}{3H_0^2}=(1-\frac{\mu}{4}) b_{\Lambda}+\alpha b_{Dust}(1+z)^{3}+\alpha b_{Warm}(1+z)^{3\mathcal{B}}\ \ \ \ \ 
\ee
\be
\n{16c}
\frac{4\rho_3 /3}{3H_0^2}= -\mu b_{\Lambda}+4\alpha b_{Dust}(1+z)^{3}+ [\mathcal{B}-(1+\mathcal{B})\mu]b_{Warm}(1+z)^{3\mathcal{B}}\ \ \ \ \ 
\ee
\end{subequations}

Coupling constants $\alpha$ and $\mu$ are related by the condition $\lambda_-=1$ that leads to the relationship $\mu=(4-4\alpha-3\mathcal{B})/(1+\mathcal{B})$ and so $\lambda_+=\mathcal{B}$. \\
Within the framework of cosmological models of accelerated expanding universes, the equations (\ref{16}) give some constraints on $\alpha$ and $\mu$. For example, from  equation (\ref{16b}), it can be seen that $\alpha$ should be positive because this density of energy must be positive at early times.   \\
Also, since $\mathcal{B} > 1$, from  equation (\ref{16a}) when $z >> 1$, it results that at least, $\mu > 0$. The upper limit for $\mu$ is given by equation (\ref{16c}) when $z \rightarrow 0$, that is,  $\mu \leq (4\alpha b_{Dust}+\mathcal{B} b_{Warm})/(b_{\Lambda}+(1+\mathcal{B} )b_{Warm})$. As $\alpha > 0$ and $\mu > 0$ is $\mathcal{B}= (7-4\alpha)/(3+\mu)-1<4/3 $. 
The case $\alpha=0$ is clearly prohibited in this work because it produces a constant behavior of effective vacuum $\rho_2$.   Therefore, the interaction affecting the vacuum cannot be identically zero here.  
 Note that the model is valid until a near future, $z_{lim}<0$, for which $\rho_3(z_{lim})=0$ that is  $z_{lim}\sim -1+ [\mu b_{\Lambda}/([\mathcal{B}-\mu(1+\mathcal{B})]b_{Warm})]^{1/3\mathcal{B}}$ or $z_{lim}\sim -0.0079$  for the best fit parameters.\\
The expressions for the density parameters $\Omega_i = \rho_i/\rho$ of the interacting fluids  are

\begin{subequations}
\n{17}
\be
\n{17a}
\Omega_1 = \frac{\mu b_{\Lambda}+(1-4\alpha)b_{Dust}(1+z)^{3}+ \mu (1+\mathcal{B})b_{Warm}(1+z)^{3\mathcal{B}}}{b_{\Lambda}+b_{Dust}(1+z)^{3}+ b_{Warm}(1+z)^{3\mathcal{B}}},
\ee
\be
\n{17b}
\Omega_2 =\frac{(1-\frac{\mu}{4}) b_{\Lambda}+\alpha b_{Dust}(1+z)^{3}+\alpha b_{Warm}(1+z)^{3\mathcal{B}}}{b_{\Lambda}+b_{Dust}(1+z)^{3}+ b_{Warm}(1+z)^{3\mathcal{B}}},\ \ \ 
\ee
\vskip0.3cm
\no and
\vskip0.3cm
\be
\n{17c}
\frac{4}{3}\Omega_3 = \frac{ -\mu b_{\Lambda}+4\alpha b_{Dust}(1+z)^{3}+ [\mathcal{B}-(1+\mathcal{B})\mu]b_{Warm}(1+z)^{3\mathcal{B}}}{b_{\Lambda}+b_{Dust}(1+z)^{3}+ b_{Warm}(1+z)^{3\mathcal{B}}}.
\ee
\end{subequations}
\vskip1cm

Equation (\ref{17b}) leads us to fix the value of $\alpha$, that is rather limited by the stringent bounds on the density parameter of DE, reported at recombination era and/or at Big Bang Nucleosynthesis (BBN). Also, the asymptotic value of $\Omega_2$ is $\alpha$ and must be in good agreement with the forecast of  Planck  and CMBPol experiments \cite{Calabrese:2010uf} as well as  with the upper bound provided by  the constraints on the variation in the fine structure constant (must be $\Omega_2<0.06$ \cite{Calabrese:2011nf} or even $\Omega_2<0.04$ if the constraints are performed with Cosmic Microwave Background, Large Scale Structure, Supernovae Ia and the Boomerang \cite{Doran:2006kp}). \\

Here, we could fix the asymptotic value of $\Omega_2$, $\alpha$,  and from the best fit of $\mathcal{B}$ obtain the value of $\mu$. Instead, we consider more appropriate to make an adjustment for all parameters involved, since the constant coefficients $b_i$ are really functions of the actual deceleration parameter $q_0$, the redshift of transition $z_{acc}$ and the coupling constants $\alpha$ and $\mu$,
\vskip0.5cm

\begin{subequations}
\n{18}
\be
\n{18a}
b_{\Lambda}(q_0,z_{acc},\mathcal{B}(\alpha,\mu)) = \frac{[2(1+q_0)-3\mathcal{B}](1+z_{acc})^{3}+(3\mathcal{B}-2)(1-2q_0)(1+z_{acc})^{3\mathcal{B}}}{3[-2(\mathcal{B}-1)-\mathcal{B}(1+z_{acc})^{3}+(3\mathcal{B}-2)(1+z_{acc})^{3\mathcal{B}}]}
\ee
\be
\n{18b}
b_{Dust} (q_0,z_{acc},\mathcal{B}(\alpha,\mu)) =\frac{2[(1+q_0)(3\mathcal{B}-2)(1+z_{acc})^{3\mathcal{B}}+(2(1+q_0)-3\mathcal{B})]}{3[-2(\mathcal{B}-1)-\mathcal{B}(1+z_{acc})^{3}+(3\mathcal{B}-2)(1+z_{acc})^{3\mathcal{B}}]}, 
\ee
\no and
\be
\n{18c}
b_{Warm}(q_0,z_{acc},\mathcal{B}(\alpha,\mu))=\frac{2[-(1+q_0)(1+z_{acc})^{3}+(1-2q_0)]}{3[-2(\mathcal{B}-1)-\mathcal{B}(1+z_{acc})^{3}+(3\mathcal{B}-2)(1+z_{acc})^{3\mathcal{B}}]}.
\ee
\end{subequations}

\begin{figure}[hbpt]
  \centering
  \subfigure[]{\includegraphics[height=5cm,width=0.35\textwidth]{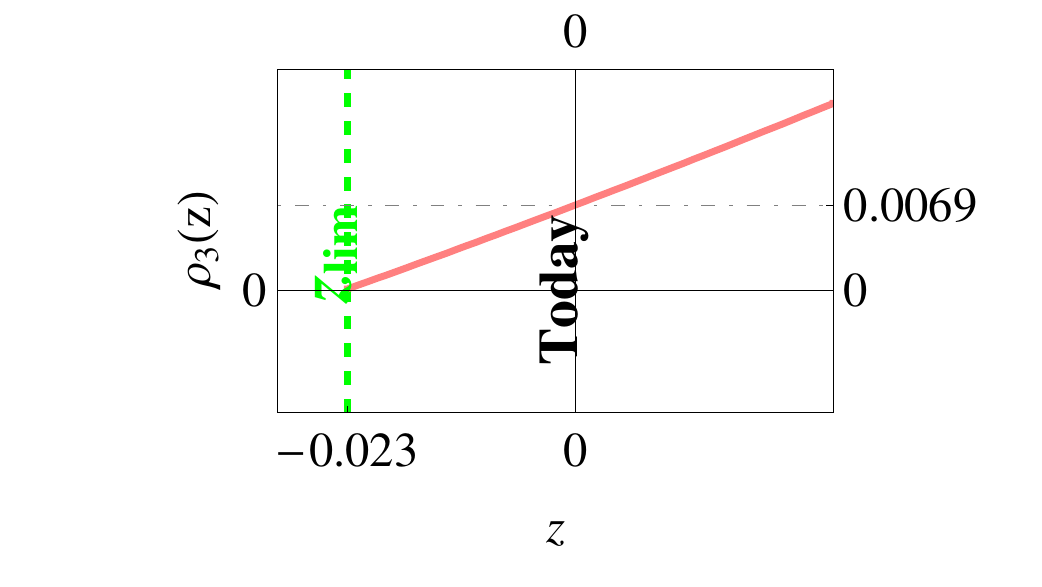}
\label{Fig:. Figura3a}}
  \subfigure[]{\includegraphics[height=6cm,width=0.55\textwidth]{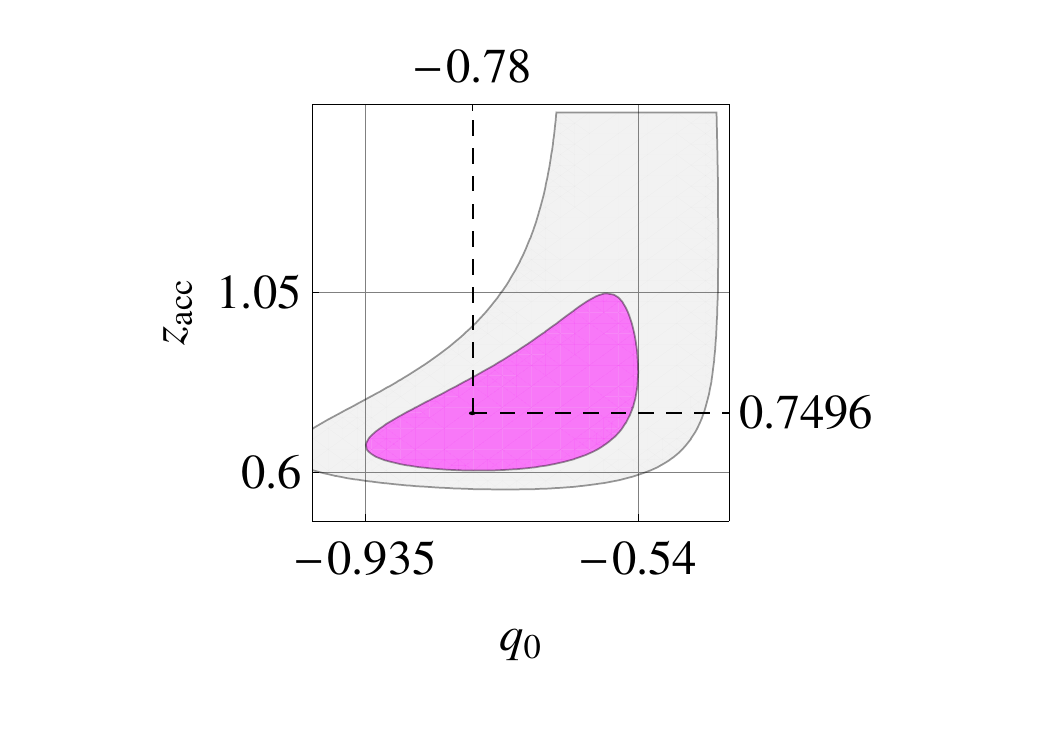}
\label{Fig:. Figura3b}} 
  \caption{\scriptsize{a) Evolution of effective density of energy for radiation and the most negative redshift  for the validity of the model, $z_{lim}=-0.023$ for the best fit parameters},\\
b)\scriptsize{Marginalized bi-dimensional confidence regions $1\sigma$ and $2\sigma$ in the parameter space $q_0$ vs $z_{acc}$ for the real interactive model. At $1\sigma$ confidence level $q_0=-0.78^{+0.54}_{-0.155}$ and $z_{acc}=0.7496^{+0.2004}_{-0.1496}$.}}
  \label{Fig:. Figura3ab}
\end{figure}

The best fit values, $H_0=74.0645$ km/sMpc, $q_0=-0.78 $, $z_{acc}=0.74959$, $\alpha=0.0000996$ and $\mu = 0.120305$ were obtained by minimizing the function $\chi^2$ that again uses (\ref{09}) but now with the constant coefficients $b_i$ expressed by (\ref{18}). The corresponding  $\chi_{dof}^2=0.962$ denotes a good fitting and in this case $z_{lim}=-0.023$ and the actual effective density of energy $\rho_3(0)=0.0069$  as it can be seen in Fig.:\ref{Fig:. Figura3a}.

\begin{figure}[hbpt]
  \centering
  \subfigure[]{\includegraphics[height=5cm,width=0.5\textwidth]{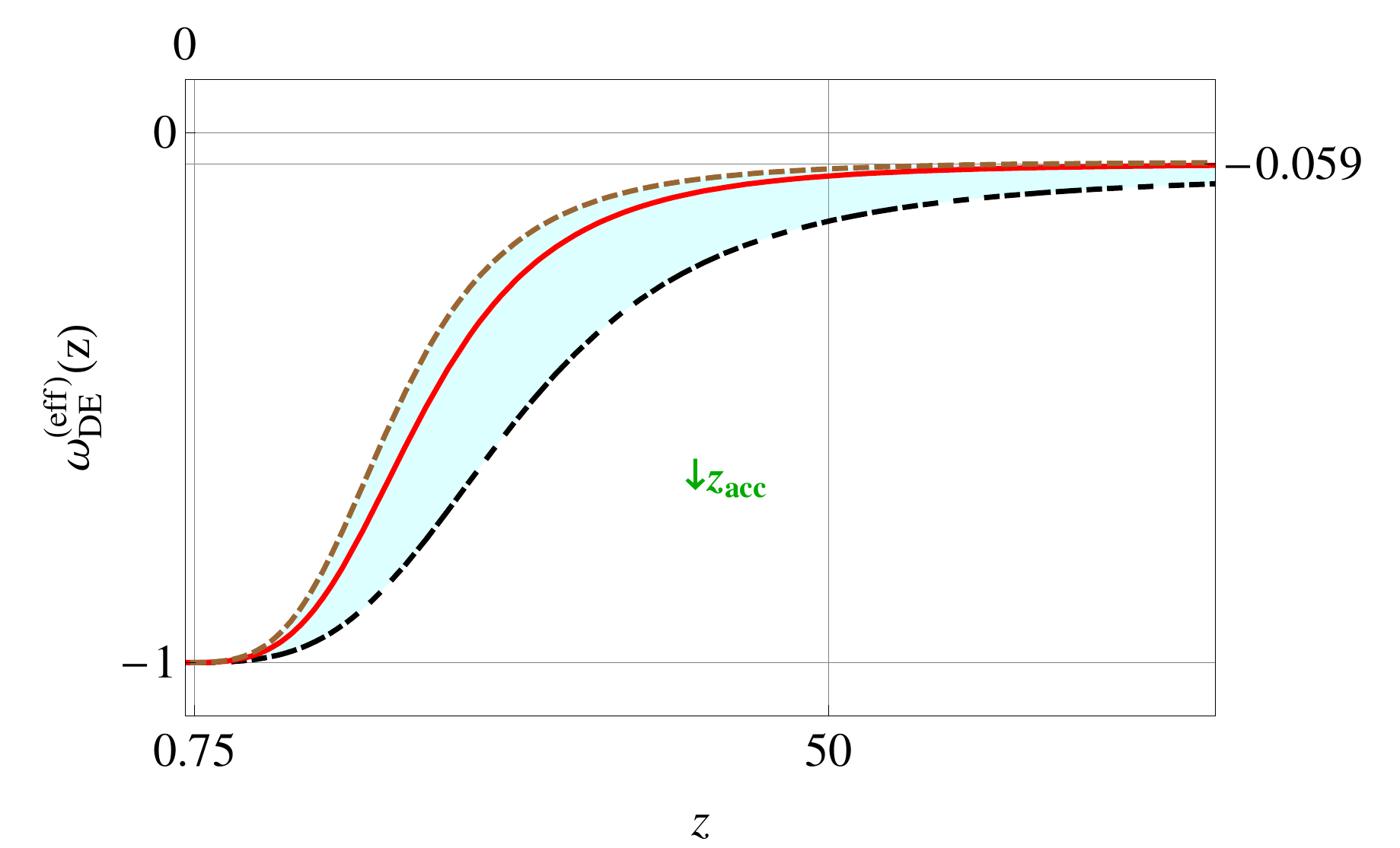}
\label{Fig:. Figura4a}}
  \subfigure[]{\includegraphics[height=6cm,width=0.45\textwidth]{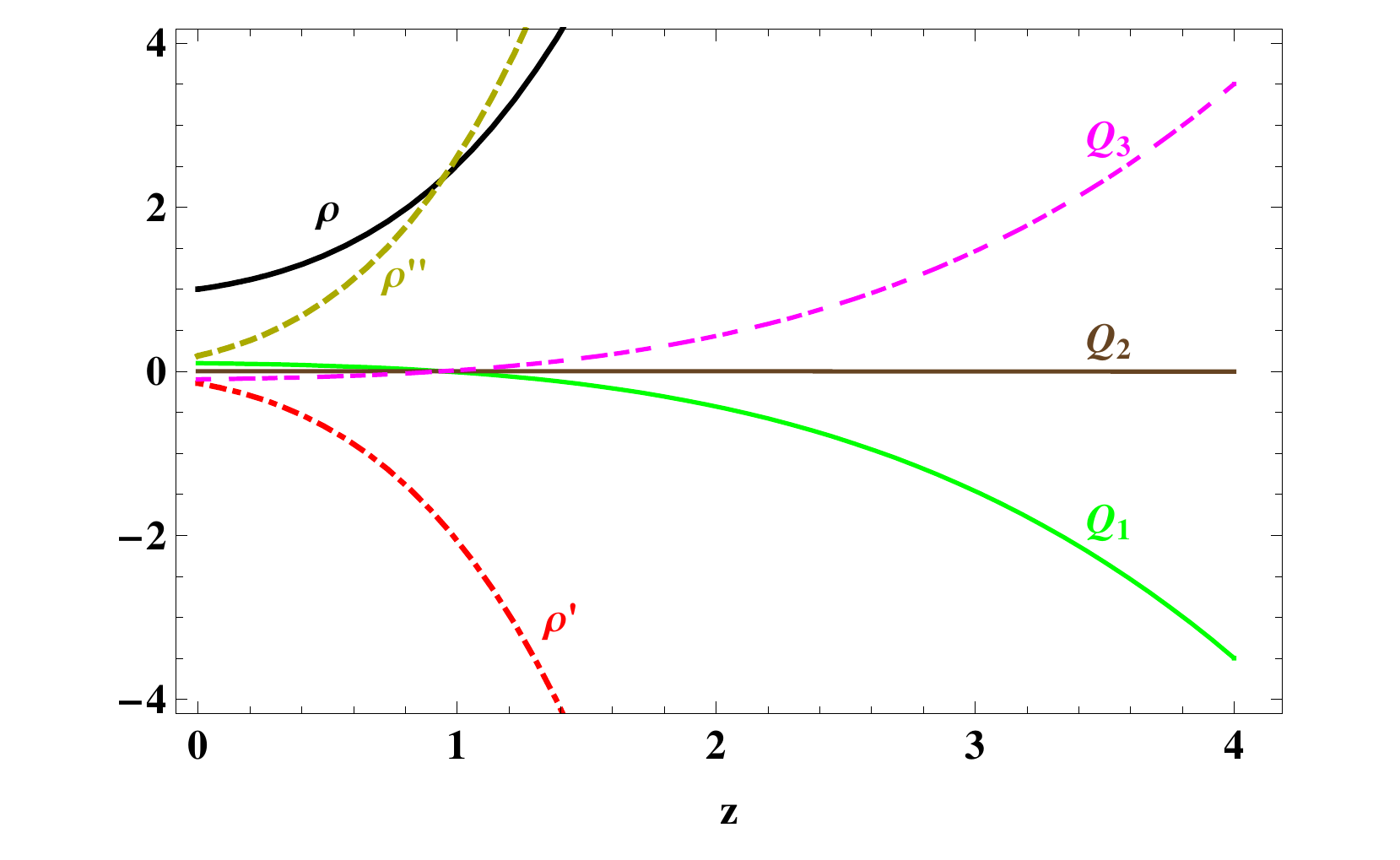}
\label{Fig:. Figura4b}} 
  \caption{
\scriptsize{ Approximated $1\sigma$ region for the evolution of the effective EoS of dark energy $\omega_{DE}^{eff}$ obtained in the real interactive model for the $1\sigma$ range of parameter $z_{acc}$.  The solid red curve corresponds to the best fit  $z_{acc}=0.749$, while values maximum and minimum of the $1\sigma$ range of $z_{acc}$, decreases (dot-dashed black curve) and increases (dashed black curve) respectively, the values of the effective EoS of dark energy},\\
b)\scriptsize{Evolution of the couplings $Q_1$ (solid  green curve), $Q_2$ (solid brown curve) and $Q_3$ (dot - dashed magenta curve), drawn in units of $3H_0^2$ for the best fit parameters, $q_0=-0.78 $, $z_{acc}=0.75$, $\alpha=10^{-4}$ and $\mu = 0.120$. Also, in units of $3H_0^2$, we show the global density  $\rho$ (solid black curve) and its derivatives first $\rho'$ (dot - dashed red curve) and second $\rho''$ (dashed darker green curve)}.}
  \label{Fig:. Figura4ab}
\end{figure}

The Fig.:\ref{Fig:. Figura3b} shows the marginalized bi-dimensional confidence regions $1\sigma$ and $2\sigma$ in the parameter space $q_0$ vs $z_{acc}$ for real interactive model. Then, the actual deceleration parameter takes values $q_0=-0.78^{+0.54}_{-0.155}$ and the values of the redshift of transition are $z_{acc}=0.7496^{+0.2004}_{-0.1496}$

The later range of variation of $z_{acc}$ was used in Fig.:\ref{Fig:. Figura4a} to depict an approximated $1\sigma$ region for the evolution of the effective EoS of dark energy $\omega_{DE}^{eff}(z)=-1-Q_2(z)/\rho_2(z)$ in the real interactive model.  The solid red curve corresponds to the best fit $z_{acc}=0.7496$, while values maximum $z_{acc}^{Max}=1.05$ and minimum $z_{acc}^{min}=0.6$ of the $1\sigma$ range of $z_{acc}$, decreases (dot - dashed black curve) and increases (dashed black curve) respectively, the values of the effective EoS of dark energy.
 
The new adjustment for all parameters confirms that the strength of interaction is weak for $Q_1$ and very weak for $Q_2$ as it can be seen in Fig.:\ref{Fig:. Figura4b}. The difference between $\rho$ (solid black curve) and $\rho''$ (dashed green curve) is very small compared with the value of $\rho'$ (dot - dashed red curve) and so the interaction $Q_1$ exhibits a much lower magnitude than $Q_2$ at all redshifts.

\begin{figure}[hbpt]
  \centering
  \subfigure[]{\includegraphics[height=3.5cm,width=0.3\textwidth]{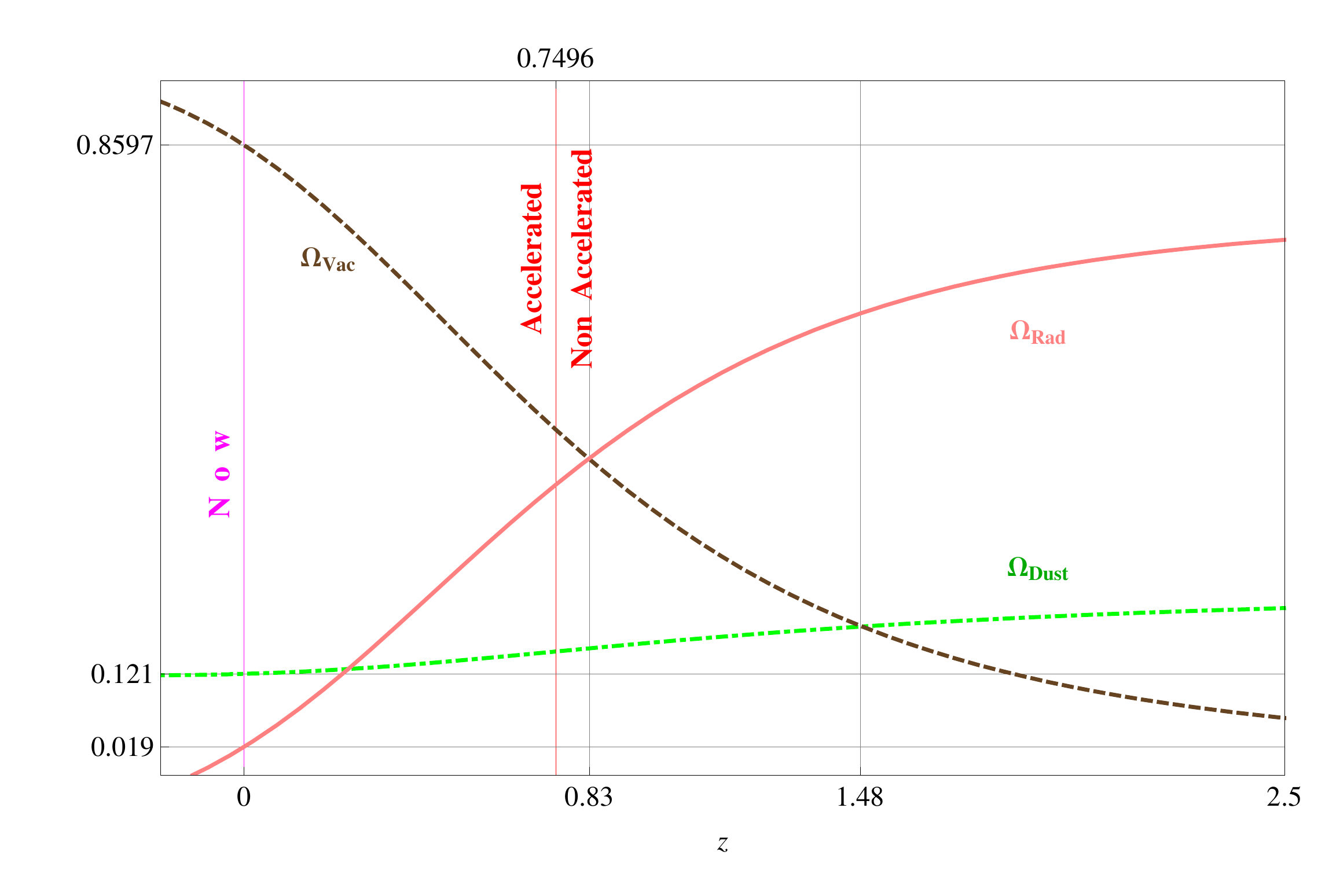}
\label{Fig:. Figura5a}}
  \subfigure[]{\includegraphics[height=3.5cm,width=0.35\textwidth]{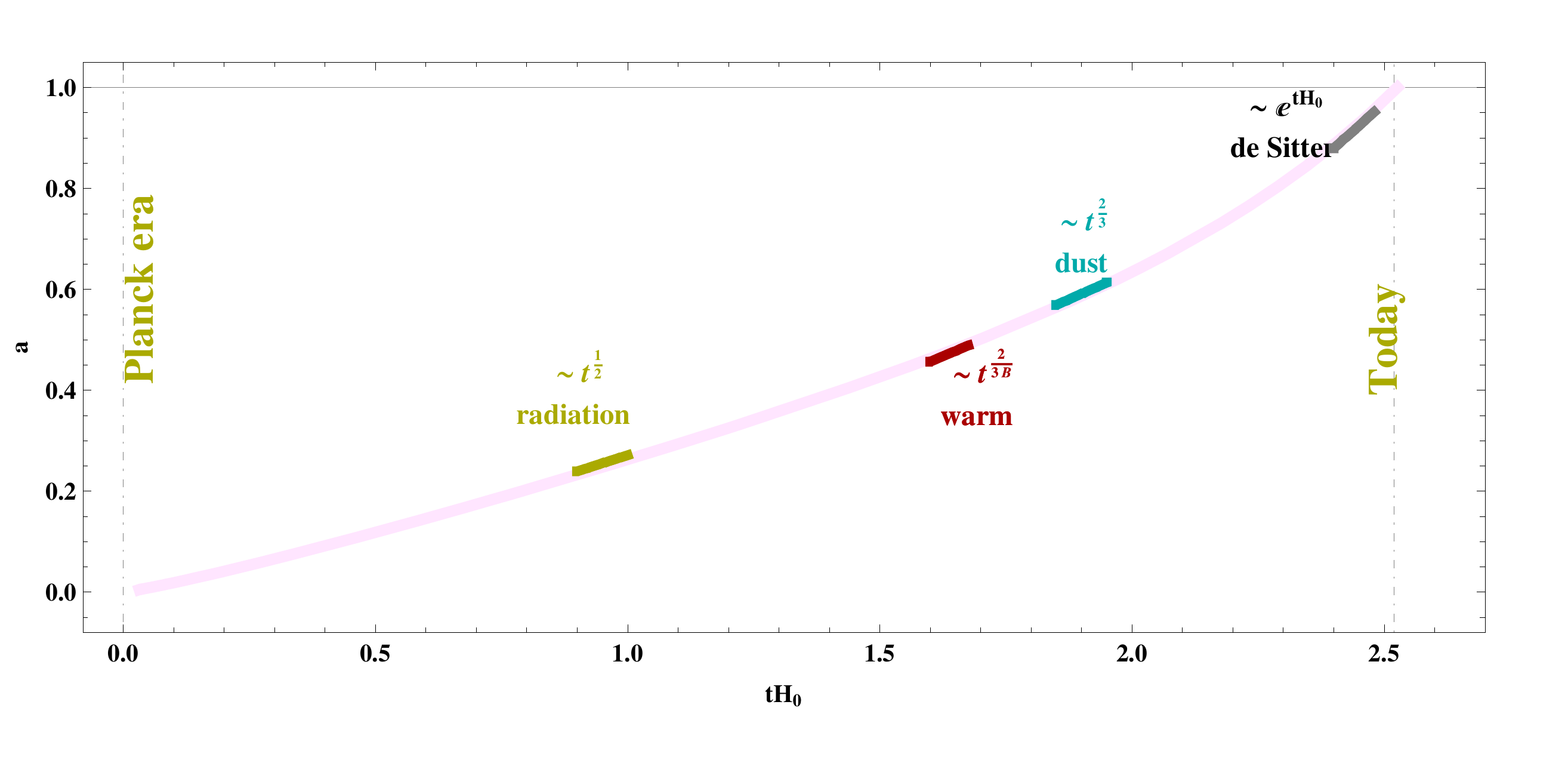}
\label{Fig:. Figura5b}}
  \subfigure[]{\includegraphics[height=3.5cm,width=0.25\textwidth]{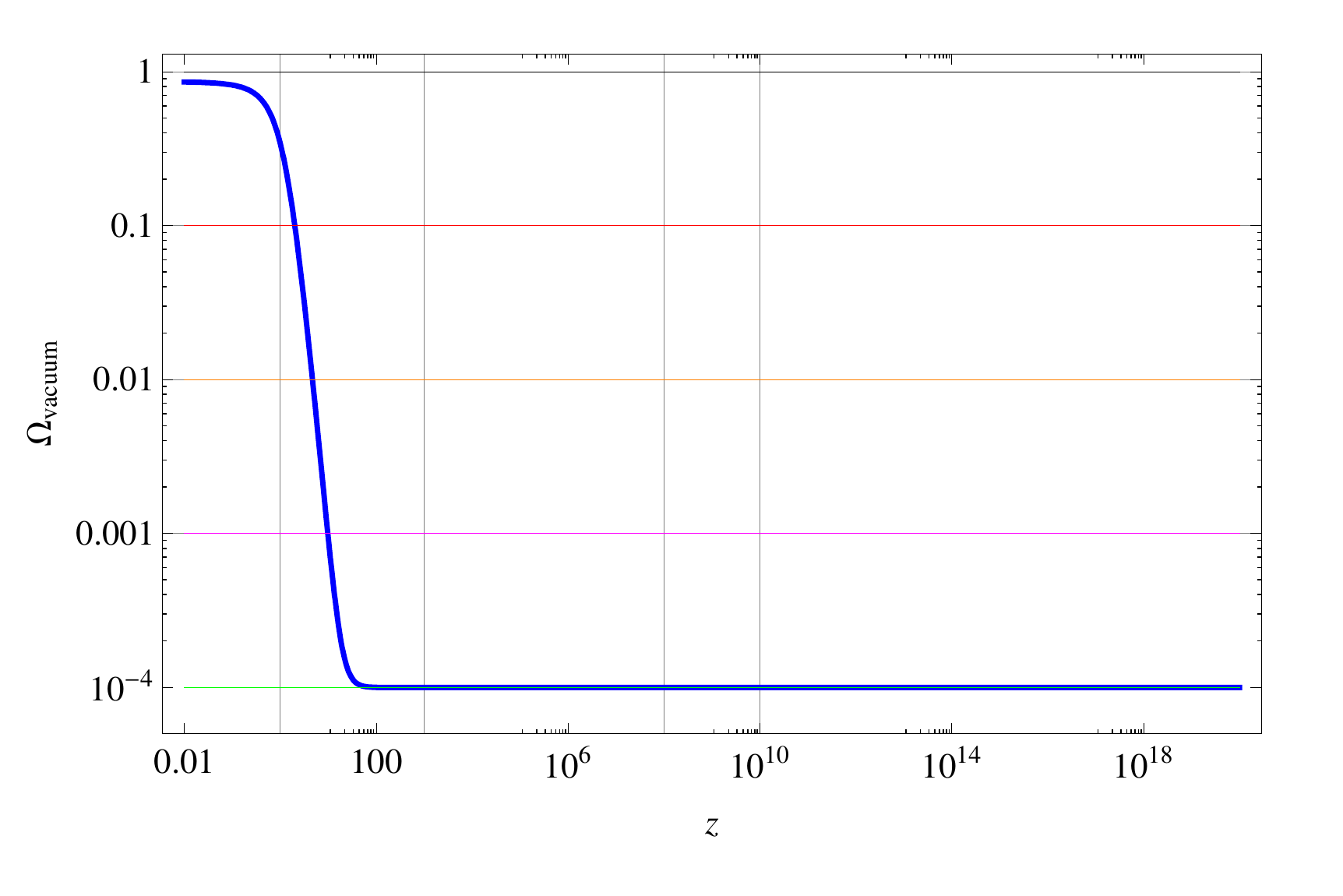}
\label{Fig:. Figura5c}} 
  \caption{\scriptsize{a) Density parameters of the real interacting fluids using the best fit parameters $q_0=-0.78$, $z_{acc}=0.75$, $\alpha=0.0000996$ and $\mu = 0.120$},\\
b)\scriptsize{parametric plot for the factor of scale $a(t)$ in $H_0^{-1}$ units with the cosmological  dominance eras  using the best fit parameters},\\
c)\scriptsize{Details of the evolution of the density parameter of vacuum energy at early times.}}
  \label{Fig:. Figura5abc}
\end{figure}

The evolution of density parameters (\ref{17}) with these best fit values of all model parameters, are shown in Fig.:\ref{Fig:. Figura5a}. In panel a) we can see that there are three different stages: radiative dominance in the distant past, then  material dominance and, just before the transition of non accelerated - accelerated universe, dark energy domination. In panel b) the parametric plot of evolution of the factor of scale $a(t)$ in $H_0^{-1}$ time units shows the same cosmological dominance eras than a) plus a dominance warm era using the best fit parameters. In panel c) it can be seen that the evolution of the density parameter of vacuum in this model fits perfectly to the stringent bounds specified above. \\

\subsection{The cosmological constant problem}

There is another interesting feature of our model with respect to the evolution of vacuum energy when it is affected by the interaction $Q_2$ and can alleviate the so-called cosmological constant problem.
A characteristic feature of general relativity is that the source for the gravitational field is the entire energy-momentum tensor. The actual value of the energy matters, not just the differences between states and this behavior opens up the possibility of vacuum energy: a density of energy characteristic of empty space that it not picks out a preferred direction. 
The scalar used by Einstein trying to find a static cosmological model, called the cosmological constant $\Lambda$ is by far, the most used invariant in mimic the vacuum energy. However, we have no insight into its expected value, since it enters as an arbitrary constant. Contributions to its value come  from zero-point fluctuations, the energies of quantum fields in their vacuum state. The inputs of all modes of oscillation (with wave number $k$) of these perturbations, give a divergent result, but on the grounds that we trust the theory only up to a certain ultraviolet momentum cut-off $k_{\mathrm{max}}$, we find that the resulting density of energy is of the form $\rho_{\mathrm{vac}}\sim \hbar k_{\mathrm{max}}^4$. 
If we are sure that we can use ordinary quantum field theory all the way up to the reduced Planck scale $\overline{m}_{\mathrm{Planck}}\sim 10^{18}\mathrm{GeV}$, we expect a contribution of order $\rho_{\mathrm{vac}} \sim  (10^{18}\mathrm{GeV})^4 \sim  10^{112}\mathrm{erg/cm}^3$. 
Nevertheless, the cosmological observations imply $\rho^{\mathrm{obs}}_{\mathrm{vac}}\leq (10^{-12}\mathrm{GeV})^4\sim 10^{-8}\mathrm{erg/cm}^3$ much smaller than the naive expectation just derived and is the origin of the famous discrepancy of 120 orders of magnitude between the theoretical and observational values of the cosmological constant.  This conundrum is the ``cosmological constant problem."

\begin{figure}[hbpt]
  \centering
  \subfigure[]{\includegraphics[height=4cm,width=0.30\textwidth]{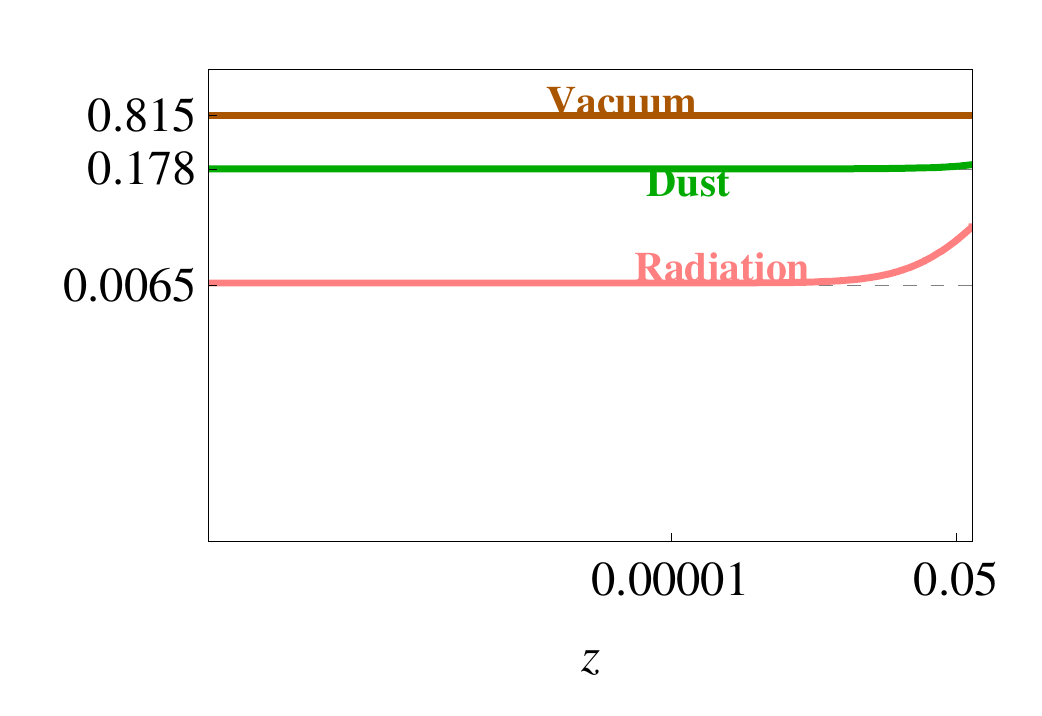}
\label{Fig:. Figura6a}}
   \subfigure[]{\includegraphics[height=3.5cm,width=0.30\textwidth]{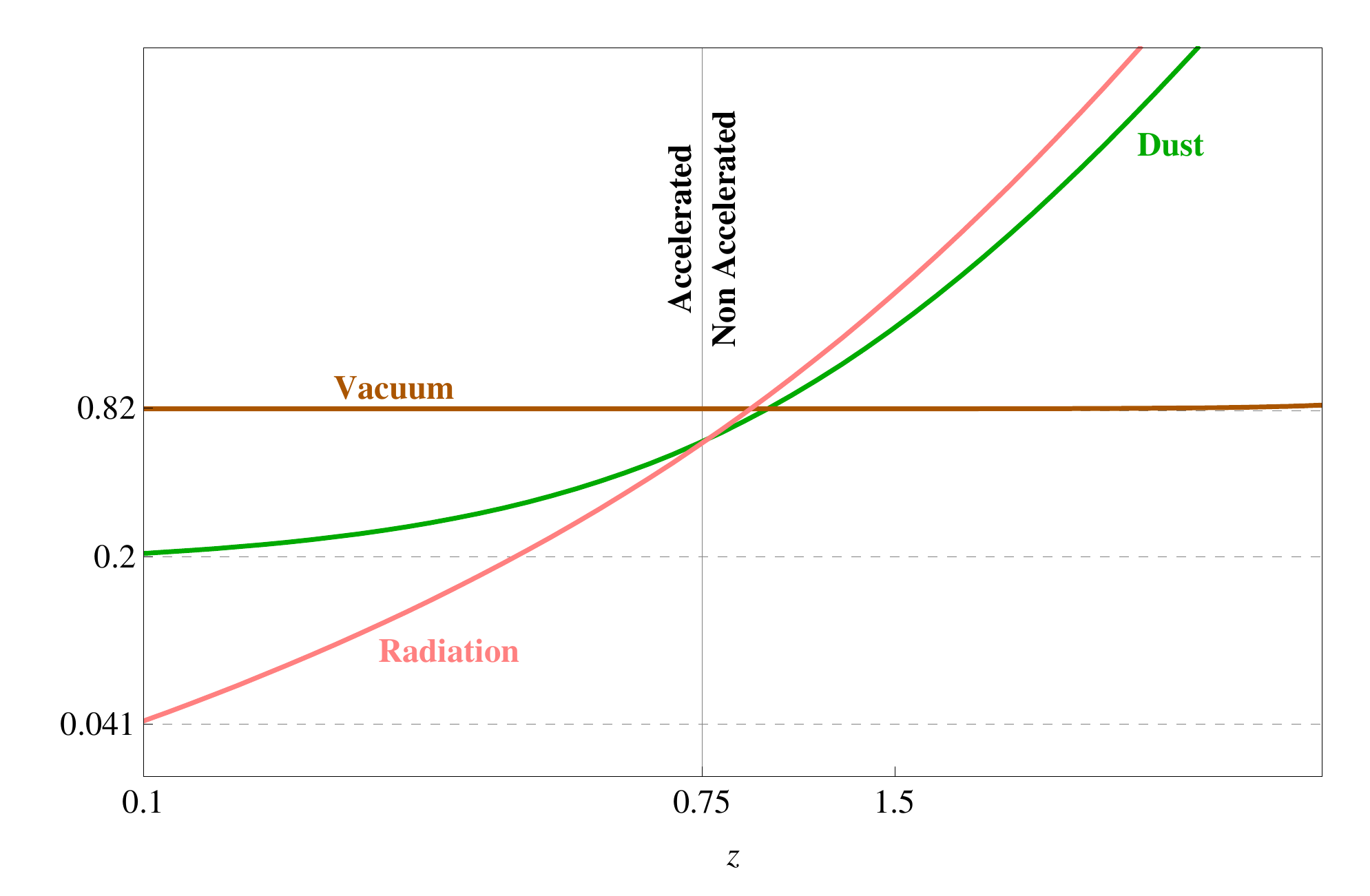}
\label{Fig:. Figura6b}}
   \subfigure[]{\includegraphics[height=4cm,width=0.30\textwidth]{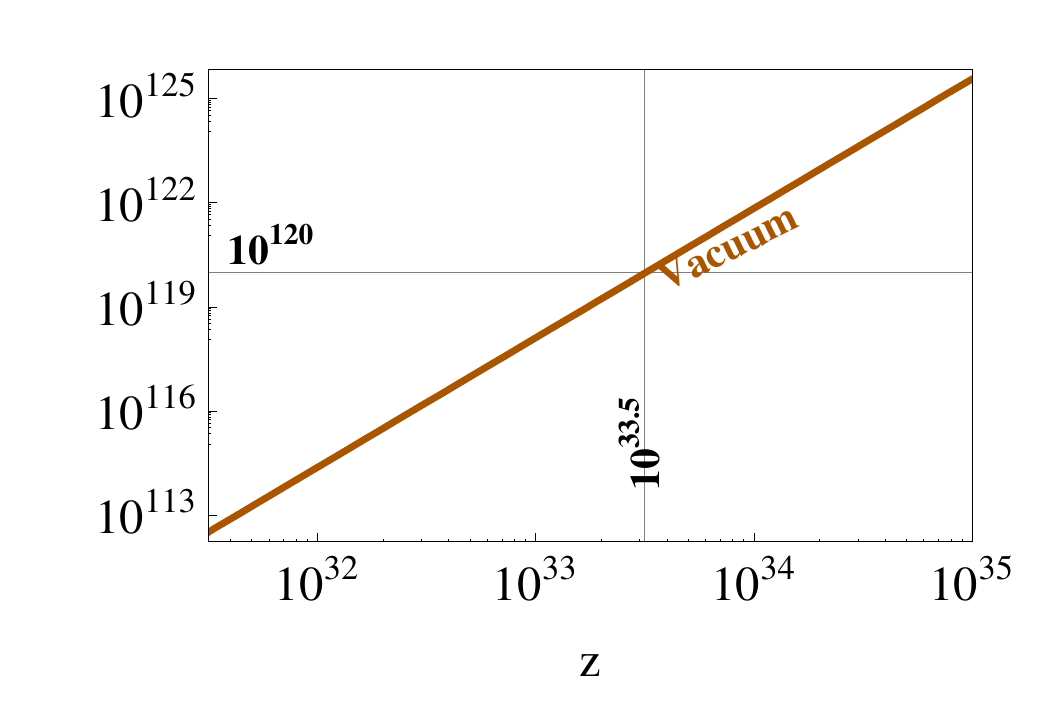}
\label{Fig:. Figura6c}}
  \caption{\scriptsize{Cosmological evolution of the interactive energy densities in ${3H_0^2}$ units. In a) and b), it can be seen the good agreement with the literature for the actual values of this fluids. In c), the interactive vacuum density of energy reaches the value predicted by the ordinary quantum field theory for the Planck energy cut off}}
  \label{Fig:. Figura6abc}
\end{figure}

In our model, the density of energy of vacuum $\rho_2$ is affected by the interaction $Q_2$ establishing a link between the value derived for very early stages of the universe, near the Planck time, and the observed value today.
From equations (\ref{17b}) we have drawn in Fig.:\ref{Fig:. Figura6a} and Fig.:\ref{Fig:. Figura6b}, the evolution of the density of energy of non relativistic matter $\rho_1$ (green curve), of the density of energy of vacuum  $\rho_2$ (brown  curve), and of relativistic matter $\rho_3$ (pink curve), in units of $3H_0^2$ for the best fit parameters $q_0=-0.78 $, $z_{acc}=0.749$, $\alpha=10^{-4}$ and $\mu = 0.120$.

Particularly, the expression (\ref{17b}) filled with the best fit parameters, allows us to obtain a contribution $10^{120}$ times larger than the present day density of energy of vacuum, when the redshift is about of $z=10^{33.5}$ as it is depicted in Fig.:\ref{Fig:. Figura6c}.

This finding shows the way to resolve, through the interaction among all components, the discrepancy between observed and theoretical values of the vacuum energy, playing the role of cosmological constant, if we can assume that $z=10^{33.5}$ is related to an age much later than the Planck time. 

To find out to what time this redshift corresponds, we need a relation t(z) between cosmological time and redshift.


\subsection{The time-redshift relation}

\begin{figure}[hbpt]
\begin{center}
\makebox[\textwidth]{\includegraphics[width=\linewidth]{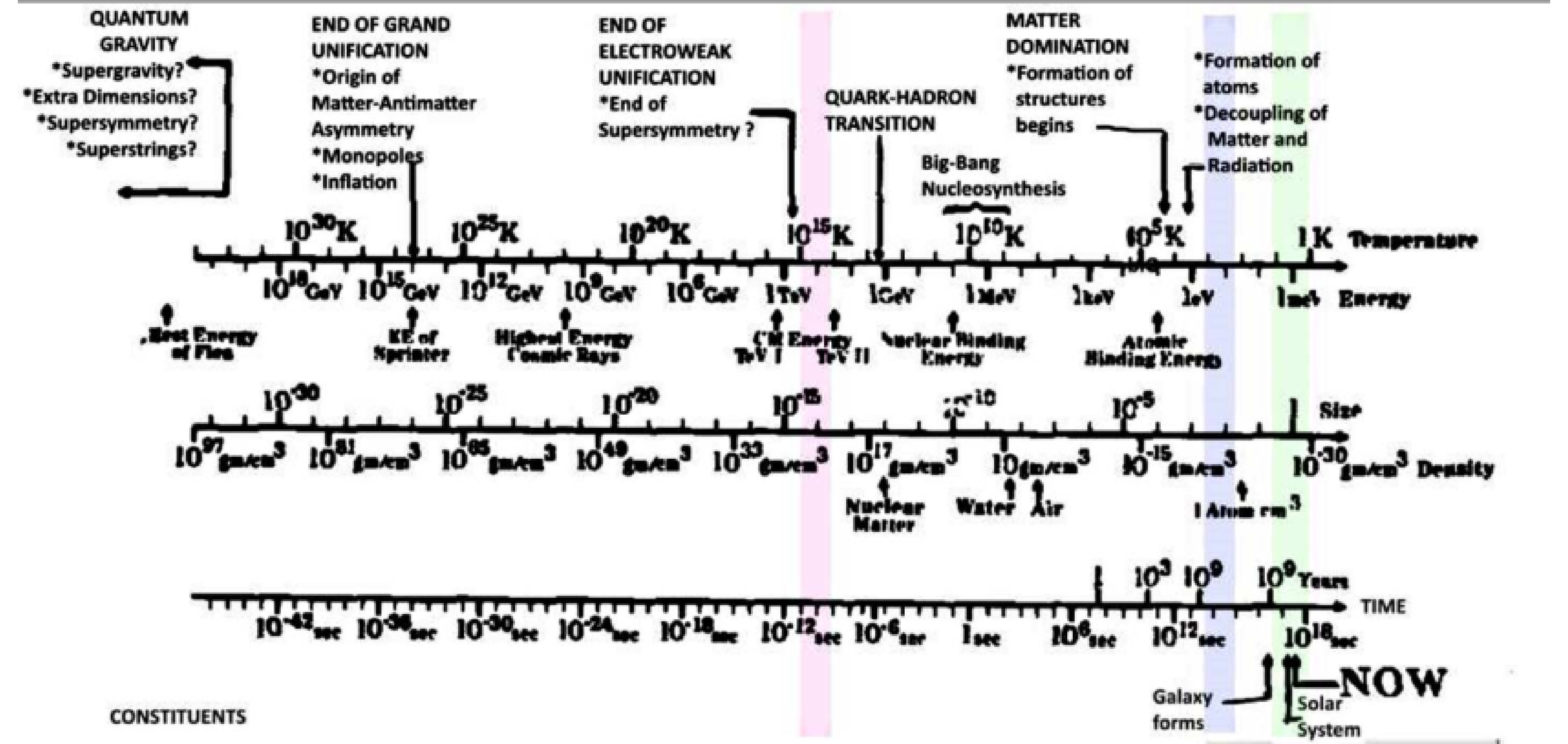}}
\caption{\scriptsize{Graphic adapted from Kolb \cite{Kolb} with our results superimposed. The lines indicate the present time (green line), the time of decoupling $z\sim 1100$ (lightblue line) and the epoch where the density of vacuum energy get the ``right" value $z\sim 10^{33.55}$ (pink line).}}
\label{Fig:. Figura 7}
\end{center}
\end{figure}

The cosmic age-redshift relation for our model reads
\be
\n{tiemp-z}
H_0 t(z)= \int_z^{\infty} \frac{dx}{(1+x)(H(z)/H_0)},
\ee
\no where $H(z)$ is taken from (\ref{09}), the time origin is set at $z = \infty$ and the time is measured in units of $H_0^{-1}$. 
For the best fit parameters, we found that the age of universe is $T=32.475 {~\rm Gyr}$ or  $T=1.024 \ 10^{18}$ sec. The value is not very close to the one reported by $WMAP-7$ year project, thus it found a $T=13.75\pm 0.13{~\rm Gyr}$ with $WMAP$ only and $T=13.75 \pm 0.11{~\rm Gyr}$ with $WMAP+BAO+H_{0}$ \cite{Jarosik:2010iu} but is still compatible with the thermal history of the Universe (see Fig.:\ref{Fig:. Figura 7}) \cite{Kolb}.  
Particularly, we situate the last scattering surface or decoupling era at time $t_{dec}=2.98\ 10^{15}$ sec.

\begin{figure}[hbpt]
  \centering
  \subfigure[]{\includegraphics[height=4.5cm,width=0.45\textwidth]{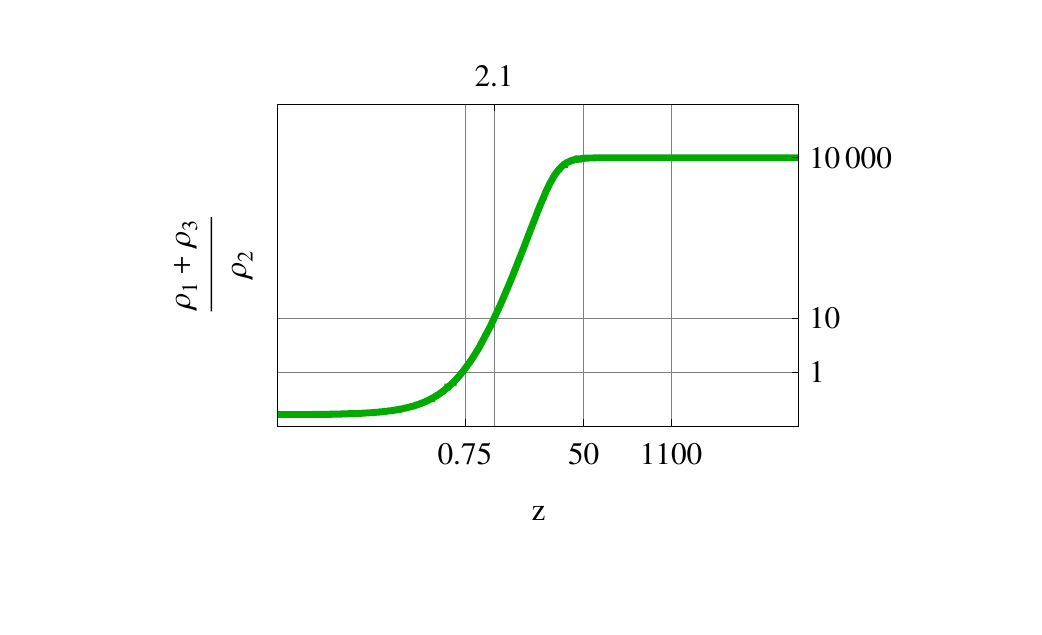}
\label{Fig:. Figura8a}}
   \subfigure[]{\includegraphics[height=5cm,width=0.45\textwidth]{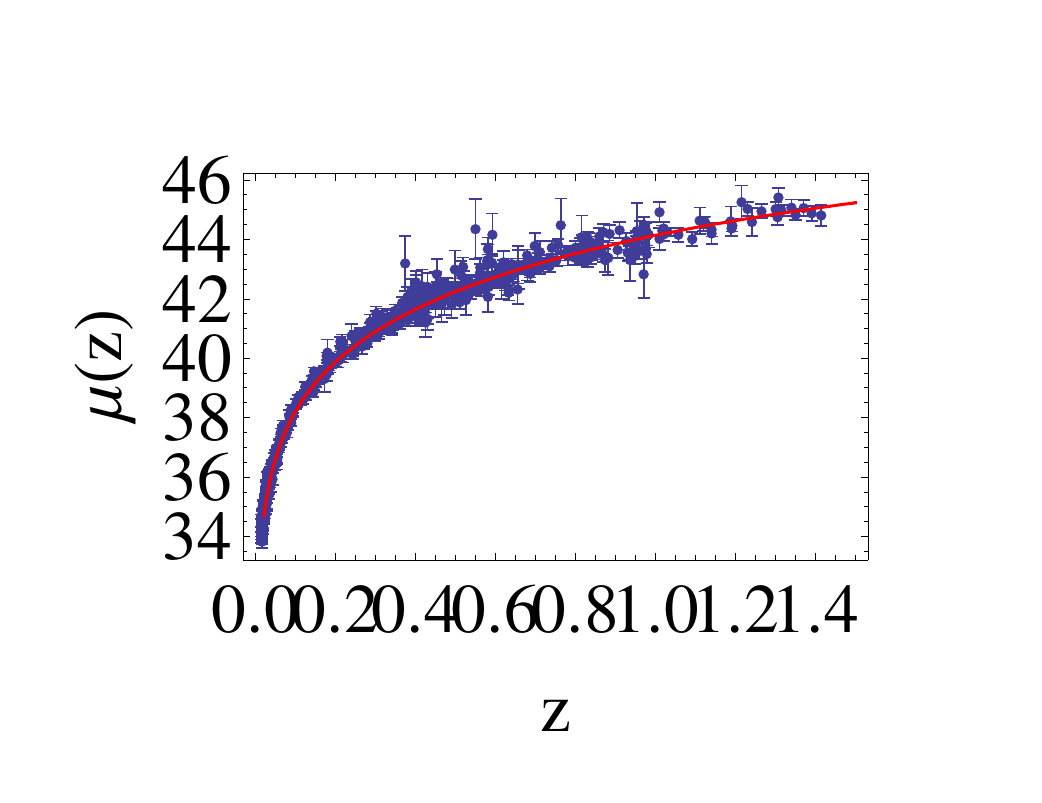}
\label{Fig:. Figura8b}}
\caption{a)\scriptsize{Evolution of the relation matter/dark energy that shows the alleviation of coincidence problem},\\
b)\scriptsize{Distance modulus for our best fit interactive model and Union2.1 data.}}
 \label{Fig:. Figura8ab}
\end{figure}

\begin{figure}[hbpt]
\begin{center}
\makebox[\textwidth]{\includegraphics[width=\linewidth]{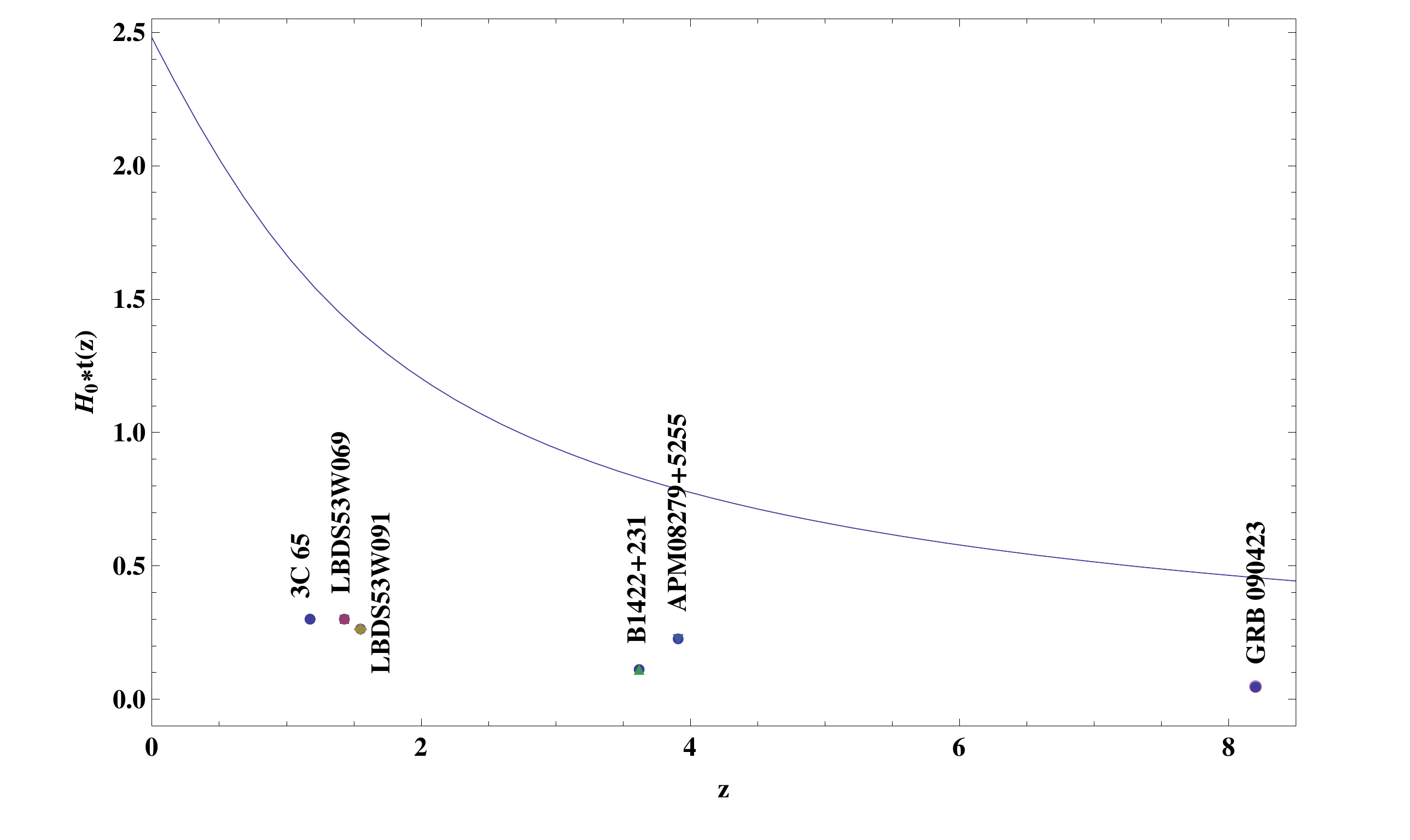}}
 \caption{\scriptsize{Cosmic age-redshift relation for the best fit parameters, in units of $H_0^{-1}$, and the old high redshift objects: radio galaxy 3C 65, galaxy LBDS 53W091, galaxy LBDS53W069, quasar APM$08279+5255$, quasar B1422+231 and the most distant known object GRB 090423.}}
\label{Fig:. Figura9}
\end{center}
\end{figure}

   With that cosmic age-redshift relation we can state that the ``cosmological constant"  was $10^{120}$ times greater than now when the universe was about $2.09\  10^{-11}$ sec, at a much later stage than the  Planck time.
Regardless of the numbers obtained, this fact highlights the feasibility of considering  interactions between non dark components to obtain a behavior very close to the model $\Lambda$CDM.



\subsection{The cosmological coincidence problem }


An important question is why the density of energy of the DE is of the same order of magnitude as the density of energy of DM even though the expansion rate of both energies is quite different. \textsl{Why are the matter and dark energy densities of precisely the same order today?}  This so called ``Cosmological Coincidence Problem'', was first formulated in Steinhardt$'$s contribution to the proceedings of a conference celebrating the 250th anniversary of Princeton University \cite{Steinhardt:1996sg}. Since then many textbooks and review papers have addressed that issue \cite{dInverno:1992gxs,Zlatev:1998tr,Steinhardt:1999nw}. Among them are those who consider the description of DE through the cosmological constant, or with a variable vacuum energy density which implies a non-gravitational interaction with DM (see, e.g., \cite{Borges:2005qs},\cite{Gomez-Valent:2014rxa} and references therein). A different line of thinking relies on anthropic considerations in which conditions for the existence of observers in an ensemble of astronomers set upper bounds on the DE density \cite{Weinberg:1987dv,Martel:1997vi,Garriga:1999bf,Lineweaver:2003px,Egan:2007ht}. On the other hand, models of interaction between DE and DM have been successful in alleviating this problem for a significant fraction of the lifetime of the Universe \cite{Amendola:1999er,Zimdahl:2001ar,Chimento:2003iea,Zimdahl:2002zb,delCampo:2008jx}. And also using the time-redshift relation $t(z)$, the interactive model can be qualified with respect to the coincidence problem, through the fraction of the age of the universe $T$ for which the ratio between dark sector densities remains around the unity. As in \cite{Forte:2013fua} such a function \textsl{quality} can be defined by 

$$\textsl{quality}(z_{coinc})= 1 - \frac{t(z_{coinc})}{T}$$
That is, the set of interactions (\ref{13})  produces a redshift of cosmic coincidence $z_{coinc}$ for which $\rho_{DM}(z_{coinc})/\rho_{DE}(z_{coinc})=1$ and \textsl{quality}
gives us a good idea of the benefits of the interaction under study respect to that issue. As it can be seen in Fig.:\ref{Fig:. Figura8a} is $z_{coinc}=0.75$ and so $\textsl{quality}=0.26$. Then during $26\%$ of the history of the universe the model satisfies the special situation of having dark densities of the same order.


\subsection{The age crisis at high redshift}


The cosmological age crisis is the well known problem of the universe being younger than its constituents (see \cite{Alcaniz:1999kr}). 
In fact, the matter-dominated FRW universe must be ruled out because its age is smaller than the ages inferred from old globular clusters. 
The age problem becomes even more serious when we consider the age of the universe at high redshift, because of some old high redshift objects (OHROs) discovered, for  instance, the $3.5$ Gyr old galaxy LBDS 53W091 at redshift $z = 1.55$ \cite{Dunlop:1996mp,Spinrad:1997md}. 
But that is not the only one. 
There are OHROs (such as the just cited LBDS 53W091 and the 4.0 Gyr old galaxy LBDS53W069 at redshift $z = 1.43$ \cite{Dunlop:1999}) that have been arranged in some models \cite{Chimento:2013se,Forte:2012ww,Chimento:2012hn}. But also, there are others, uncomfortable, that refuse to fit properly under the age curves of theoretical cosmological models proposed until today. 
For example, the 4.0 Gyr old radio galaxy 3C 65 at $z = 1.175$ \cite{Stockton:1995}, and the high redshift quasar B1422+231 at $z = 3.62$ whose best-fit age is 1.5 Gyr with a lower bound of 1.3 Gyr \cite{Yoshii:1998}. Also, the old quasar APM 08279+5255 at $z = 3.91$, whose age is estimated to be 2.0 - 3.0 Gyr \cite{Hasinger,Komossa}, is used extensively. 
Besides, and to assure the robustness of our analysis, we use the most distant known object GRB 090423, localized at $z = 8.2$ and with an estimated age of 0.63 Gyr counted from the big bang \cite{nrao:2008}.
Many authors have examined the age problem within the framework of the dark energy models, see e.g. \cite{Alcaniz:1999kr}, \cite{Friaca}-\cite{Wang:2010}, and references therein. The age problem within the context of holographic dark energy model was explored in \cite{Wei:2007} and \cite{Cui}-\cite{Wei:2010}.
The Fig.: \ref{Fig:. Figura9} shows the $t(z)$ function in units of $H_0^{-1}$ and also the old high redshift objects mentioned above, which are accommodated, all of them, under the curve of time, eliminating the age crisis, at least for these milestones.



\subsection{The effective equations of state and the statefinder diagnostic of the dark energy}

\begin{figure}[hbpt]
\begin{center}
\makebox[\textwidth]{\includegraphics[width=\linewidth]{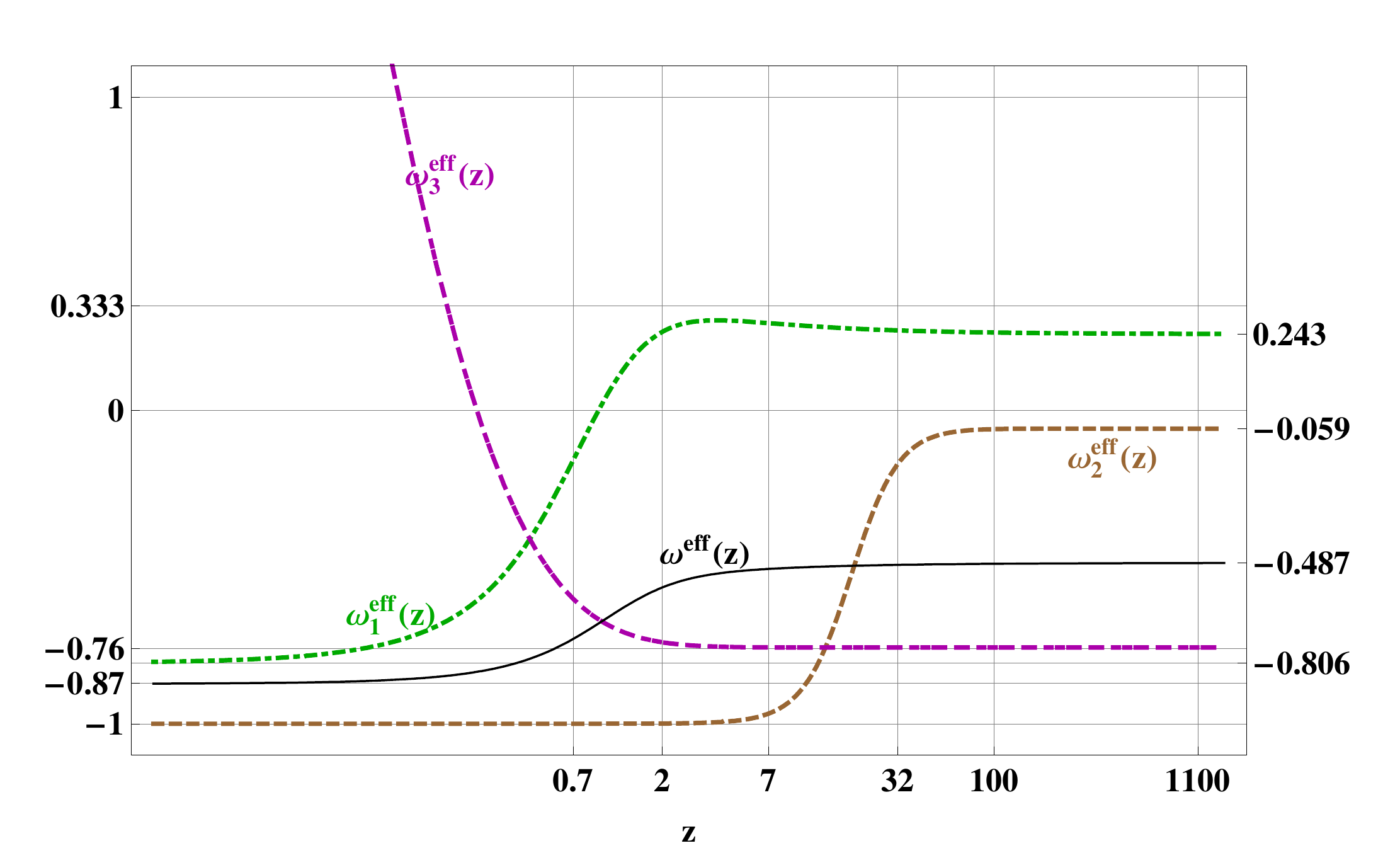}}
\caption{\scriptsize{Evolution of the effective equations of state for each interactive fluid: vacuum (dashed brown curve), baryonic matter (dot - dashed green curve) and radiation (dot - dashed magenta curve), and global effective equation of state (solid black) for the best fit parameters of the model $q_0=-0.78$, $z_{acc}=0.7496$, $\alpha=0.0000996$ and $\mu = 0.120$.}}
\label{Fig:. Figura 10}
\end{center}
\end{figure}

Although in the $\Lambda$CDM model the acceleration stage begins before the $\Lambda$DE-DM equality  while here it occurs afterwards, both models agree in the $z_{acc} = 0.75$ and show a very similar adjustment in relation to their distance modulus curves with respect to Union 2.1 data \cite{Sanchez:2009ka},\cite{Melia:2012zy}. In Fig.: \ref{Fig:. Figura8b} it can be seen this very good concordance through the drawing of the distance modulus $\mu(z)=\mu_0+5\log(D_L(z))$ where $D_L(z)$ is the luminosity distance $D_L(z)=(1 + z)\int_0^z dx \frac{H_0}{H(x)}$ and $\mu_0=42.38 -5 \log(H_0/100)$. The interesting properties of the model also arise from the effective equations of state for the individual fluids and the global one.
They are defined by the equations ${\omega_{eff}}_i(z) = \omega_{i} - Q_i(z)/\rho_i(z)$, and the expressions arise by combining equations (\ref{02}), (\ref{07}) and (\ref{09}), when the interactive energy densities $\rho_i(z,H_0,q_0,z_{acc},\alpha,\mu)$ are obtained from the equations (\ref{10}) by expressing the coefficients $b_i$ as functions of the parameters $(q_0,z_{acc},\alpha,\mu)$ with (\ref{10}). The Fig.:\ref{Fig:. Figura 10} displays these effective equations of state ${\omega_{eff}}_i(z)$ for each interactive fluid and also the global effective equation of state ${\omega_{eff}}(z)={\omega_{eff}}_1(z)\Omega_1(z)+{\omega_{eff}}_2(z)\Omega_2(z)+{\omega_{eff}}_3(z)\Omega_3(z)$ for the best fit parameters of the model,  $q_0=-0.78$, $z_{acc}=0.7496$, $\alpha=0.0000996$ and $\mu = 0.120$.

In these Fig.:\ref{Fig:. Figura 10} it can be seen that their individual effective  behaviors ${\omega_{eff}}_i(z)$ differ substantially from the constant characteristic behaviors of the fluids involved, $\omega_i= -1, 0, 1/3 $ due to the effects of the couplings,  even at present time, when it can be said that such interactions are not strong. At early times we cannot speak of an era of ``purely" radiative dominance due mainly, to the interplay with the dust energy, as it is shown in Fig.:\ref{Fig:. Figura 7}. Instead, the effective fluid acts as a mixture with the asymptotic value ${\omega_{eff}}^{asymp}=0.243$, an EoS closer to radiative 1/3 but clearly affected by the interaction with matter, until $z \sim 100$. From this redshift, the EoS of vacuum energy goes down to its identification value $-1$, passing the dust value around $z \sim 32$.  With respect to the divergence shown by the radiation Eos, we can say that is caused by the interaction, which destroys the radiation as far as its density of energy  has an order of magnitude of $ 10^{-5} $, in units of $3H_0^2$,  when as $z \rightarrow 0$. So, at late times, the radiation fluid has a very small density parameter and contributes almost nothing to the overall state equation, which shows the interplay between vacuum and matter passing the dust-like behavior, before the  matter itself and reaching its present value $ \omega_{eff}(0)=-0.794$. This result is closed to the values $\omega(0)=-1.12^{+0.42}_{-0.43}$ consigned in \cite{Jarosik:2010iu} for WMAP.
\vskip0.5cm

        The equation of state is not a fundamental property of dark energy models because of certain ambiguity on its definition that, since we assume the Einstein interpretation of gravitational filed equations \cite{Sahni:2002fz} and a spatially flat FLRW metric, has no place in this work. 
However, it is interesting to consider the use of geometrical variables (as $H_0$ and $q_0$) when describing the properties of dark energy, for example, we include the study of the jerk parameter (the third order contribution in the expansion for kinematic luminosity distance in terms of the redshift z)  $j=-\frac{\stackrel{\cdots}{a}}{H^3 a}$, in order to compare with some simple kinematic models for the cosmic expansion based on specific parameterizations for $q(z)$ and a constant jerk parameter 
\cite{Guimaraes:2009mp}. 

In terms of redshift the jerk is written as
\be
\n{jerk}
j(z)= -\left( 1 - \frac{(1+z)\frac{d\rho(z)}{dz}}{\rho(z)} + \frac{(1+z)^2}{2}\frac{\frac{d^2\rho(z)}{dz^2}}{\rho(z)}\right),
\ee
\no and so, for the best fit values  $b_{\Lambda}=0.83$, $b_{Dust}=0.05$ and $\mathcal{B}=1.242$, and the equations (\ref{03}), (\ref{04}), (\ref{09})  and (\ref{10}),  we get $j(0)=-1.13647$.

The realistic kinematic models considered by Guimaraes et all 
\cite{Guimaraes:2009mp}, at $1\sigma$ confidence limits imply the ranges of values: $q_0\in [-0.96,-0.46]$, $j_0 \in [-3.2,-0.3]$ and $z_{acc}\in [0.36, 0.84]$, whereas the $\Lambda$CDM  predictions are  $q_0 = -0.57 \pm 0.04$, $j_0 = -1$ and $z_{acc} = 0.71 \pm 0.08$.  Our best fit value model is compatible with this data.

Really, there is a whole hierarchy of geometrical parameters $A_n$, arising from the Taylor expansion of the scale factor around the present time $t_0$
\be
\n{Taylor}
(1+z)^{-1}=\frac{a(t)}{a_0}= 1 + \sum_{n=1}^{\infty} \frac{A_n(t_0)}{n!}\left( H_0(t - t_0)\right)^n
\ee
\no where 
\be
\n{Ta}
A_n = \frac{a^{(n)}}{aH^n}    \qquad  n \in N,
\ee
\no and $a^{(n)}$ is the nth derivative of the scale factor with respect to time \cite{Sahni:2002fz}.  Of course, $A_1 = 1$ and the first are already known with different letters of the alphabet: $q=-A_2$ is the deceleration parameter, $j=-A_3$ is the jerk, $s=-A_4$ is the snap, etc.\cite{Visser:2003vq,Capozziello:2008qc,Dunajski:2008tg}.

\begin{figure}[hbpt]
  \centering
  \subfigure[]{\includegraphics[height=5cm,width=0.35\textwidth]{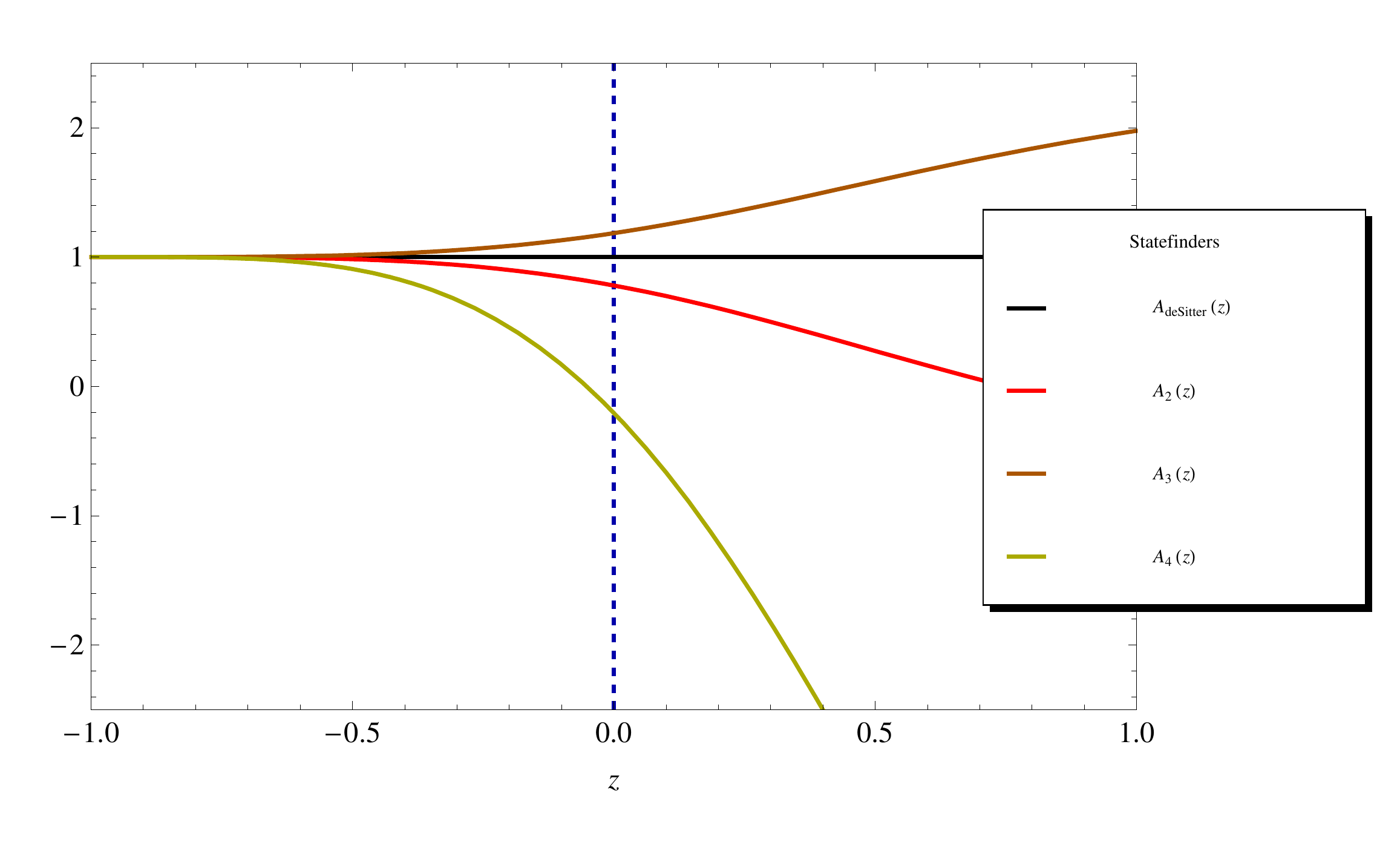}
\label{Fig:. Figura11a}}
   \subfigure[]{\includegraphics[height=6cm,width=0.60\textwidth]{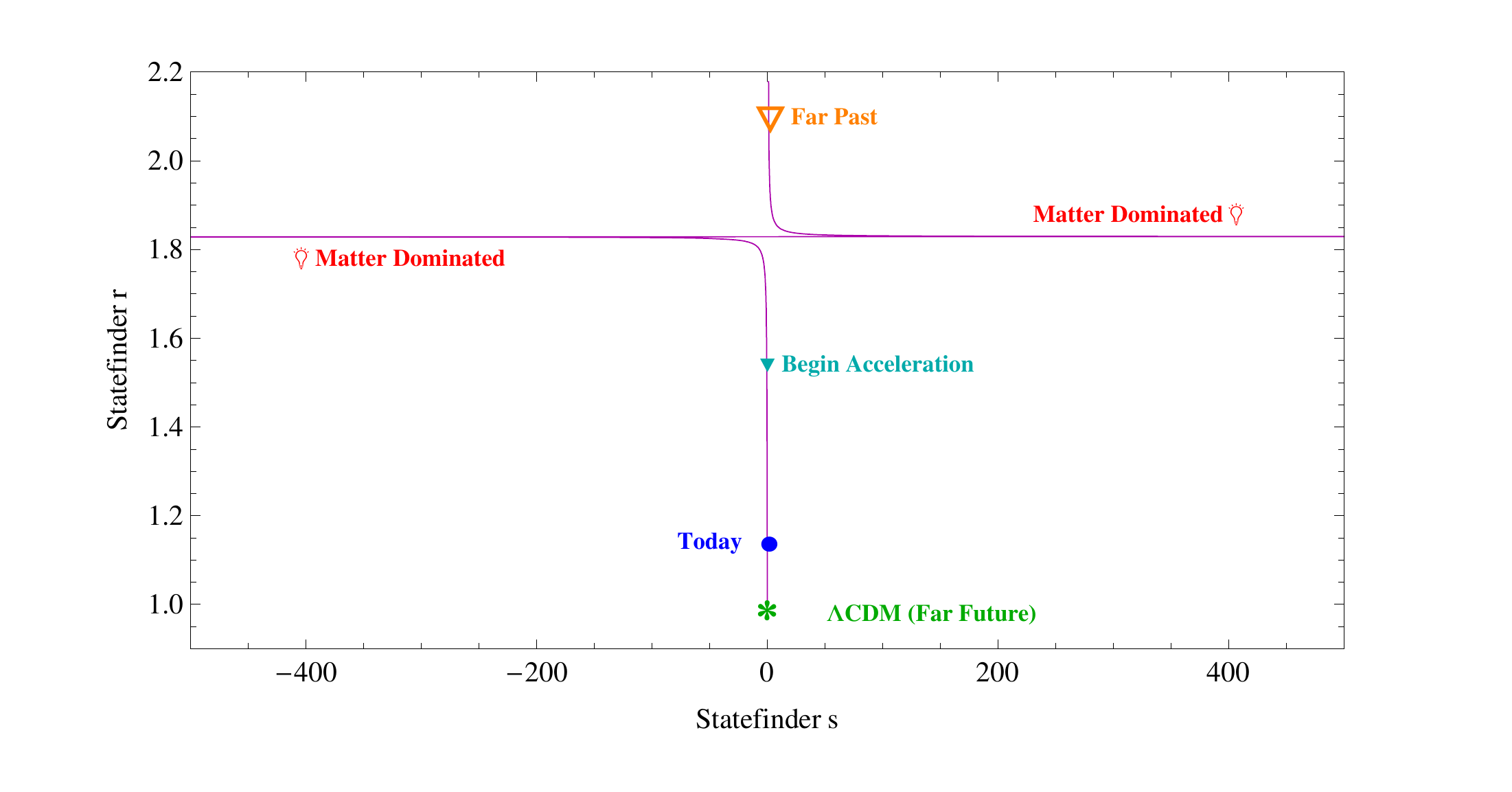}
\label{Fig:. Figura11b}}
     \caption{a)\scriptsize{Evolution of the statefinders coefficients for the interactive model.},
b)\scriptsize{Evolution of the real interactive model depicted through the statefinders par (s,r). The curve r(s) evolve in the time interval corresponds to $z \in [-1,8.5]$, that is from $ 10^{17}$ seconds after de Big Bang to the far future.}}
  \label{Fig:. Figura11ab}
\end{figure}

In Fig.:\ref{Fig:. Figura11a} we show the evolution for the first four parameters $A_n$ describing our interactive model and also, the evolution of the de Sitter model parameters, for which all of them are constant and equal to 1.

The natural next step to characterize the properties of dark energy is the Statefinder par $\{r,s\}$ \cite{Sahni:2002fz}
\be
\n{Statef}
r = \frac{\stackrel{\cdots}{a}}{H^3 a},  \qquad s = \frac{r - 1}{3(q - 1/2)}.
\ee
In Fig.:\ref{Fig:. Figura11b} we show the evolution of our interactive model in the phase-space (r,s). It can be seen a matter dominated stage, the beginning of acceleration and the future de Sitter type behavior or $\Lambda$CDM fixed point.


\section{Conclusions}

Our model begins with a universe that contains mainly energy of vacuum and where the generation of relativistic and non-relativistic particles alters the global energy density  causing the "ignition" of the interactions $Q_1$ and $Q_2$. The interactive system proposed by us is formally equivalent to a system of three self-preserved fluids that correspond to vacuum energy, baryonic matter and warm dark matter. The parameters of the equivalent system are adjusted with the observational data from the Hubble function resulting as best fit the values $H_0=74 \mathrm{km/sMpc}$ for the actual Hubble parameter, $b_{\Lambda}=\Omega_{\Lambda} = 0.844$, $b_{Dust}=\Omega_{bar}=0.047$ for the density parameters and $\mathcal{B}=1+p_{Warm}/b_{Warm}=1.243$ for the warm pressure, with a good value per degree of freedom $\chi^2_{dof}=0.992$.
 
The real model consists of three effective fluids that interact with each other:\\
a) an effective component of non-relativistic matter that includes not only the baryonic part but also the part associated with dark matter that seems to be generated purely and exclusively by interaction,\\ b) an effective component of vacuum energy, which at times of the order of $10^{-10}$ sec. ($z= 10^{32}$) has a density of energy which is compatible with the density theoretically calculated  with the Planck energy as cut off and whose density parameter respects the dimensions suggested by the physics of recombination and the  Big Bang nucleosynthesis (BBN), $\Omega_{\Lambda}<0.05$, \\
c) an effective component of relativistic matter, whose current density parameter, of the order of $10^{-5}$, agrees with the values considered in the literature and that determines the end of the validity of the model when it is canceled in the  near future $z_{lim}=-0.023$.

The  density parameters of the equivalent model can be written as functions of the current Hubble function $H_0$, the current deceleration parameter $q_0$, the redshift of transition to the accelerated stage $z_{acc}$, the constant state equation of the warm component $\mathcal{B}$ and the coupling constants $\alpha$ and $\mu $. Then, a complementary adjustment of the model with the data of the Hubble function let us to obtain the best fit values $H_0=74.0645 km/sMpc$ $q_0=0.78$, $z_{acc}=0.74959$, $\alpha = 0.0000996$ and $\mu = 0.120305$ with a goodness of adjustment of $\chi^2_{dof}= 0.962$.

The time-redshift relationship of the model $t(z)$ allows to locate it temporarily and show it in good concordance with descriptions such as the thermal history of the Kolb universe: the age of our universe is located in the same band of $10^{18}$ sec.($z= 10^{32}$) and the same thing happens with the surface of last dispersion around $10^{15}$ sec. Similarly, $10^{-10}$ sec. after the Big Bang, we are in a scenario in which it is permissible to use ordinary quantum field theory to calculate the vacuum energy.

 The temporal curve corresponding to the model allows to correctly accommodate all the oldest known stellar objects (at least up to $ z=8.2 $) solving in principle the paradox of a universe with inhabitants older than its own history (Crisis of the Age).

The curve of the scale factor obtained in parametric form exhibits a period of radiative dominance at early times, followed by a period of material dominance (warm firstly and then cold matter), to end with an accelerated universe at present.
The effective state equations for each interactive fluid have a markedly different behavior with respect to the bare constants that identify each of them: at early times, all behave as if they corresponded to a perfect mix with $\omega_i= 0.242$; at late times, the only one that retains its identity is that of vacuum energy, while that of radiation diverges when its density of energy almost disappears.
The effective density of the model shows the transition to the final accelerated universe but not a crossing of the phantom barrier.

Finally, we studied the model in the comparative scheme of the statefinders, finding that it is included within the realistic models described by Guimaraes.


\end{document}